\pdfoutput=1
\documentclass[sn-basic,Numbered]{sn-jnl}

\usepackage{graphicx}%
\usepackage{multirow}%
\usepackage{amsmath,amssymb,amsfonts}%
\usepackage{amsthm}%
\usepackage{mathrsfs}%
\usepackage[title]{appendix}%
\usepackage{xcolor}%
\usepackage{textcomp}%
\usepackage{manyfoot}%
\usepackage{booktabs}%
\usepackage{algorithm}%
\usepackage{algorithmicx}%
\usepackage{algpseudocode}%
\usepackage{listings}%
\usepackage{siunitx}%
\usepackage{pdfpages}

\theoremstyle{thmstyleone}%
\theoremstyle{thmstyletwo}%

\theoremstyle{thmstylethree}%

\begin{document}

\title[]{Single-Chip Silicon Photonic Processor for Analog Optical and Microwave Signals}


\author*[1,2]{\fnm{Hong} \sur{Deng}}\email{hong.deng@ugent.be}

\author[1,2]{\fnm{Jing} \sur{Zhang}}\email{jingzhan.Zhang@ugent.be}

\author[1,2]{\fnm{Emadreza} \sur{Soltanian}}\email{emadreza.soltanian@ugent.be}
\author[1,2]{\fnm{Xiangfeng} \sur{Chen}}\email{xiangfeng.chen@ugent.be}
\author[1,2]{\fnm{Chao} \sur{Pang}}\email{chao.pang@ugent.be}
\author[3]{\fnm{Nicolas} \sur{Vaissiere}}\email{nicolas.vaissiere@3-5lab.fr}
\author[3]{\fnm{Delphine} \sur{Neel}}\email{delphine.neel@3-5lab.fr}
\author[3]{\fnm{Joan} \sur{Ramirez}}\email{joan.ramirez@3-5lab.fr}
\author[3]{\fnm{Jean} \sur{Decobert}}\email{jean.decobert@3-5lab.fr}
\author[4]{\fnm{Nishant} \sur{Singh}}\email{nishant.singh@UGent.be}
\author[4]{\fnm{Guy} \sur{Torfs}}\email{guy.torfs@ugent.be}
\author[1,2]{\fnm{Gunther} \sur{Roelkens}}\email{gunther.roelkens@ugent.be}
\author*[1,2]{\fnm{Wim} \sur{Bogaerts}}\email{wim.bogaerts@ugent.be}

\affil*[1]{\orgdiv{Photonics Research Group, Department of Information Technology, },\orgname{Ghent University - imec}, \orgaddress{\city{Ghent}, \country{Belgium}}}

\affil[2]{Center for Nano- and Biophotonics (NB Photonics), Ghent University, \orgaddress{\city{Ghent}, \country{Belgium}}}

\affil[3]{\orgdiv{III-V Lab, } \orgaddress{\city{Palaiseau}, \country{France}}}

\affil[4]{\orgdiv{IDLab, Department of Information Technology}, \orgname{ Ghent University - imec}, \orgaddress{\city{Ghent}, \country{Belgium}}}

\abstract{The explosion of data volume in communications, AI training, and cloud computing requires efficient  data handling, which is typically stored as digital electrical information and transmitted as wireless radio frequency (RF) signals or light waves  in optical fibres. Today's communications systems mostly treat the RF and optical signals separately, which results in unnecessary conversion losses and increased cost. In this work, we report the first fully on-chip signal processor for high-speed RF and optical signals based on a silicon photonic circuit. Our chip is capable of both generation and detection of analog electrical and optical signals, and can program a user-defined filter response in both domains. The single silicon photonic chip integrates all essential components like modulators, optical filters, and photodetectors, as well as tunable lasers enabled by transfer-printed Indium Phosphide (InP) optical amplifiers. The system's configuration is locally programmed through thermo-optic phase shifters and monitored by photodetectors. We demonstrate our chip's capabilities with different combinations of RF and optical signal processing functions, including optical and RF signal generation and filtering. This represents a key step towards compact microwave photonic systems for future wireless communication and sensing applications.}


\maketitle

\section{Introduction}\label{sec1}


Today's increasingly interconnected and data-driven society relies on always-accessible high-speed data connections. Communication networks are evolving towards an increasingly granular infrastructure, powered by fixed-line optical fibers and high-bandwidth wireless access networks. To distribute the digital data, systems use separate devices to generate and process optical signals and wireless radio frequency (RF) signals, which results in unnecessary conversion and propagation losses, as well as increased cost. We need a more versatile approach to effectively handle the signals in both the RF and the optical domain.

Silicon photonic circuits have emerged as a widely deployed solution for optical signal transmission \cite{Siew2021ReviewDevelopment}, manipulating light on the surface of a chip using optical waveguides to transport signals, high-speed modulators to encode information on a lightwave, and germanium photodetectors(PDs) to convert these signals back to the electrical domain. The signals can be filtered with waveguide-based interferometers and coupled off-chip to optical fibres. Supported by the high-precision manufacturing infrastructure of electronics, there is a clear trend towards increasing circuit complexity and programmability \cite{Bogaerts2022ProgrammableCircuits}. Through the integration of phase tuners, the paths of light can be electronically configured to implement wavelength/frequency filters with tunable pass bands, arbitrary linear transformers, and convolution calculators  \cite{Wang2022ProgrammableMZI,Macho-Ortiz2021OpticalTransformations,Shen2017DeepCircuits}, which prove to be invaluable for quantum information processing, deep learning, and LiDAR systems\cite{Sparrow2018SimulatingPhotonics, Wang2018MultidimensionalOptics,Ashtiani2022AnClassification, Sun2013Large-scaleArray}. 

Because optical signals can span a massive bandwidth, photonics is used to generate, manipulate, transport, and measure high-speed RF signals. This field is commonly known as Microwave Photonics (MWP) \cite{Capmany2007MicrowaveWorlds,Yao2009MicrowavePhotonics}. Enabled by fast electro-optic modulators, tunable optical filters and low-noise PDs, photonic circuits can perform diverse complex RF signal processing functions \cite{Marpaung2019IntegratedPhotonics, Perez-Lopez2020MultipurposeCircuits,Liu2016AProcessor, Fandino2016AFilter, Guo2021VersatileShaper}. Current demonstrations are based on InP platforms \cite{Liu2016AProcessor, Fandino2016AFilter}, silicon photonic platforms \cite{Guo2021VersatileShaper} and SiN platforms \cite{ Perez-Lopez2020MultipurposeCircuits}. On-chip lasers and optical amplifiers can only be natively supported on InP platforms. No demonstration with a natively integrated laser source has been presented on other platforms. 

Silicon has an indirect bandgap, and therefore the integration of III-V materials, such as GaAs and InP, has been extensively explored \cite{Roelkens2007III-V/SiBonding,Matsumoto2019Hybrid-integrationBonding, Zhang2019III-V-on-SiMicro-transfer-printing, Liu2018PhotonicSilicon}. Transfer printing has emerged as a promising method that optimizes the utilization of costly III-V epitaxial substrates without modifications to the fabrication process of silicon photonics.

In this work, we report the first single-chip signal processor built in silicon photonics technology, capable of implementing programmable operations on both optical and RF signals, and converting between the two domains. The chip contains two widely tunable lasers, a reconfigurable modulator, optical switches, a fully tunable Mach-Zehnder interferometer (MZI) loaded with 4 independent ring resonators, and two high speed PDs. The silicon photonic chip is fabricated in imec's iSiPP50G process, supplemented with micro-transfer-printed optical amplifiers. The system’s configuration is locally programmed through thermo-optic phase shifters and monitored by on-chip tap PDs.

\begin{figure}
    \centering
    \includegraphics[width=\linewidth]{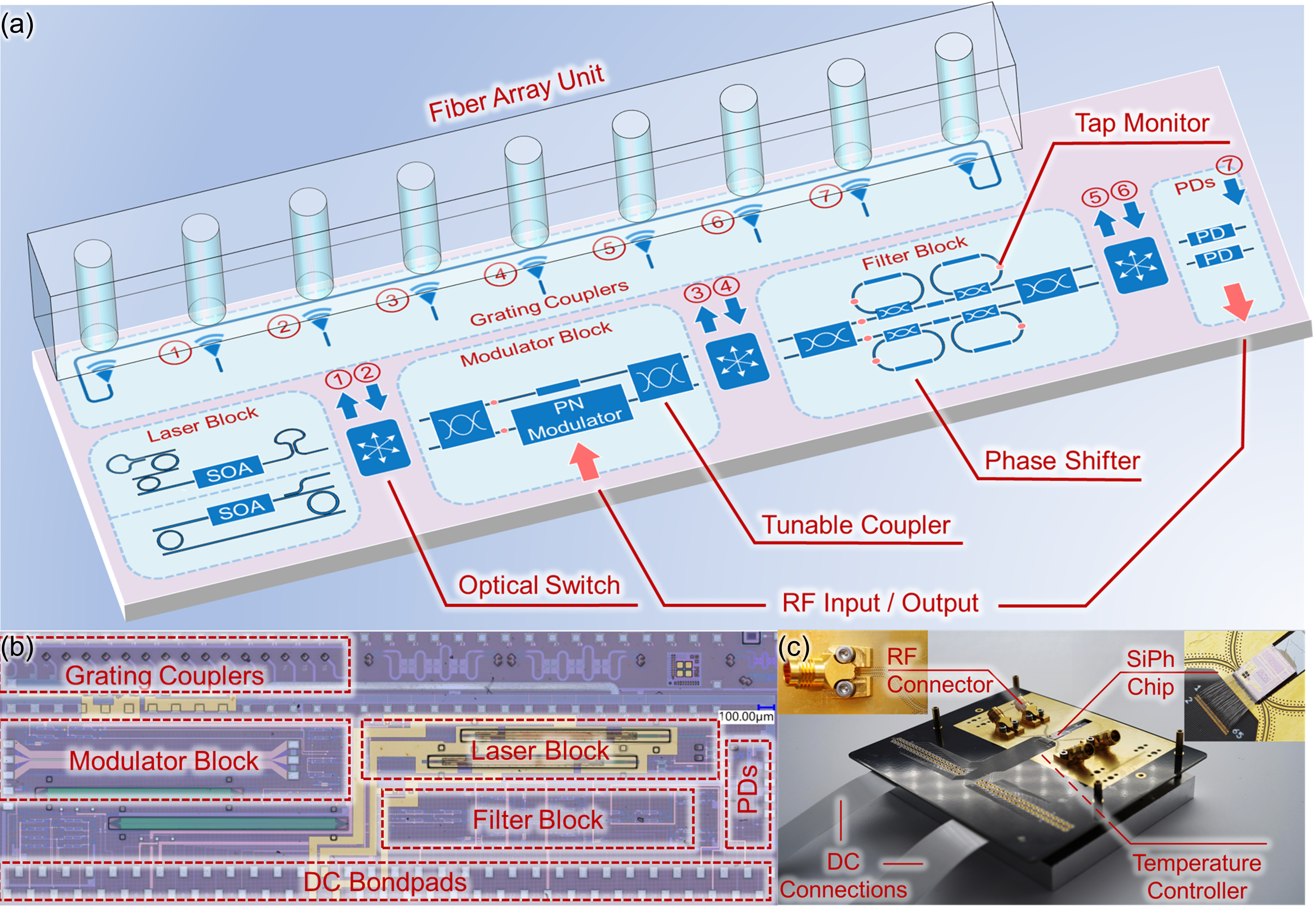}
    \caption{(a) Schematic block diagram of the silicon photonic signal processor. (b) Microscope image of the fabricated chip; (c) A packaged demonstrator with wirebonded controls and microwave connectors (before fibre attachment).}
    \label{fig:1}
\end{figure}
 
\section{Results}\label{sec2}

Our signal processor consists of four main subcircuit blocks, which are the tunable laser  (with 2 lasers), the reconfigurable modulator, the reconfigurable optical filter and the high-speed PD block, shown in Fig.~\ref{fig:1}a. Light generated by the laser source(s) is guided into the modulator circuit, which can be configured for either phase modulation or intensity modulation (details in supplementary information S2.3). The light, now carrying an RF payload signal, feeds into the optical filter consisting of a fully tunable four-ring-loaded MZI with tap monitors. The filtered light signal is then converted back into the RF domain by the high-speed PD. The four functional blocks are connected by optical switches, making it possible to inject or extract light at each joint, allowing  processor configurations for both optical and RF signal processing, or cascading of  processors through fiber links. The optical switches  are implemented as single-stage or double-stage MZIs (details in supplementary information S2.2) \cite{Wang2020TolerantCircuit}. A fabricated chip is shown in Fig.~\ref{fig:1}b, and a  packaged device (before fiber attachment) is shown in Fig.~\ref{fig:1}c. The entire photonic circuit has 15 optical fibre ports, 1 RF input and 2 RF outputs, 52 heater-based optical phase shifters, and 8 tap monitors, and it has the capability of fully reconfiguring the optical and electrical response and the conversions between the domains, and it can also function as a tunable laser source or a tunable RF source. 
\\

\begin{figure}
    \centering
    \includegraphics[width=\linewidth]{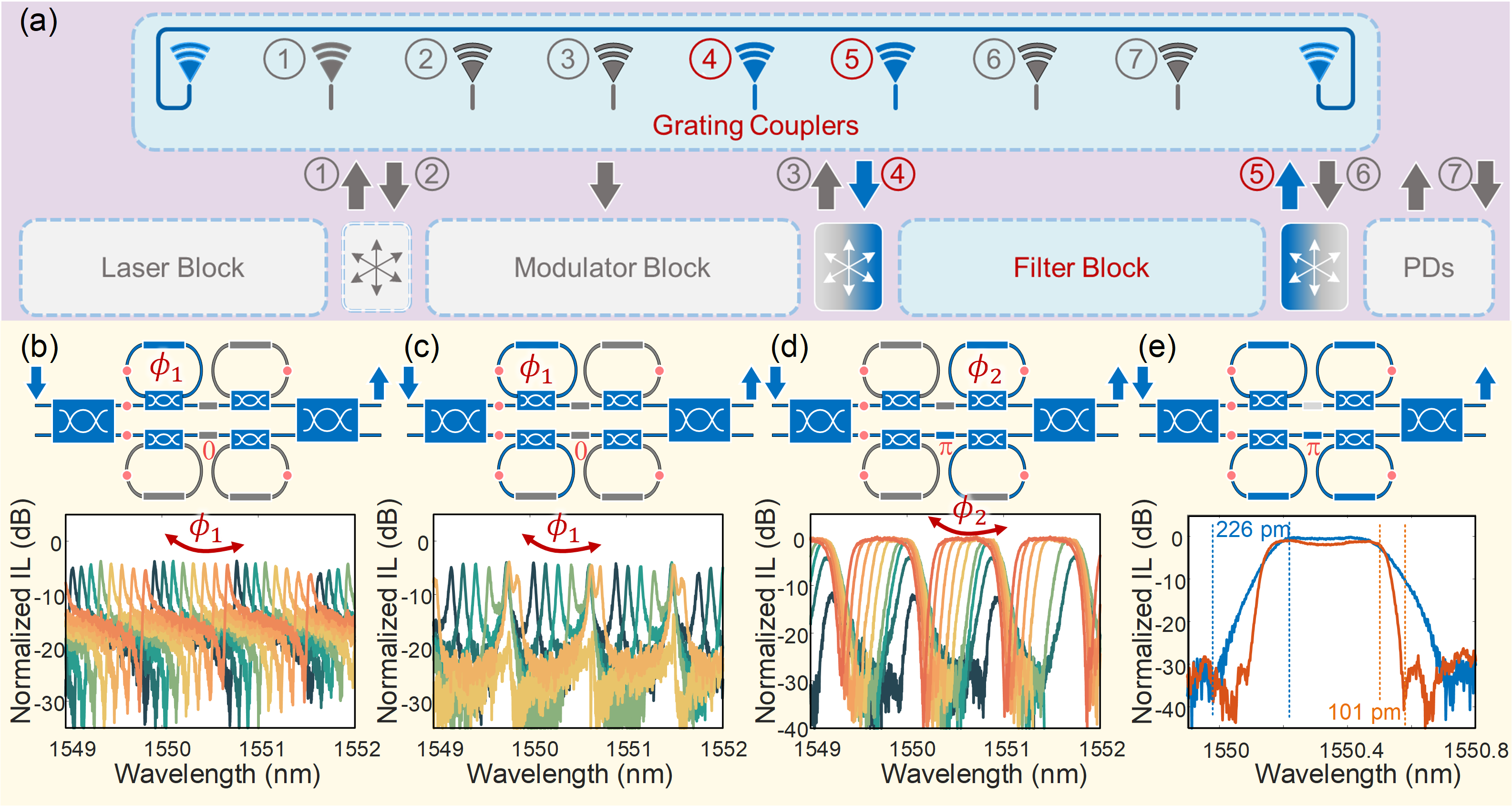}
    \caption{(a) The signal processor configuration for optical filtering. (b) Single bandpass filtering configuration and response; (c) Double bandpass filtering configuration and response; (d) Second-order Chebyshev Type II bandpass filtering configuration and response; (e) Forth-order Chebyshev Type II bandpass filtering configuration and response.}
    \label{fig:2}
\end{figure}

\noindent\textbf{Optical-to-optical (O/O) response.}  Figure~\ref{fig:2} pictures  the signal processor configured as an optical linear filter: the light signal is guided through the MZI loaded with four independently tunable ring resonators (two in each arm). Such ring-loaded MZIs have can act as an auto-regression/moving average (ARMA) filter \cite{Madsen1999OpticalAnalysis, Wang2022ProgrammableMZI,Fandino2017AFilter, Luo2010HighInterleaver}. We implemented all couplers in the filter as tunable MZIs, as shown in Fig~\ref{fig:2}b-e. With the built-in tap monitors, the coupling status of each ring can be calibrated locally. The tuning procedures are explained in the supplementary information~S2.4. The optical responses shown in Fig.~\ref{fig:2}b-e are measured with an optical vector analyzer (LUNA OVA5000), and normalized to grating coupler transmission envelope. If the phase difference in the MZI is zero, the rings will introduce sharp pass bands as shown in Fig.~\ref{fig:2}b-c. The fabricated rings show a \qty{3}{dB} bandwidth of \qty{35}{pm} when critically coupled. If the MZI phase difference is set to $\pi$, an overcoupled ring pair can form typical Chebyshev Type II filters with a flat passband. The measurements show passbands with  $\sim$\qty{0.5}{dB} ripple in passbands and \qty{30}{dB} stopband rejection (Fig.~\ref{fig:2}d). By then tuning the second ring pair to the roll-off points, a higher-order Chebyshev Type II filter response is obtained, improving the roll-off bandwidth  from \qty{226}{pm} to \qty{101}{pm} (Fig.~\ref{fig:2}e). More details and examples of optical filtering are described in the supplementary information S2.4. 
\\

\begin{figure}[!t]
    \centering
    \includegraphics[width=\linewidth]{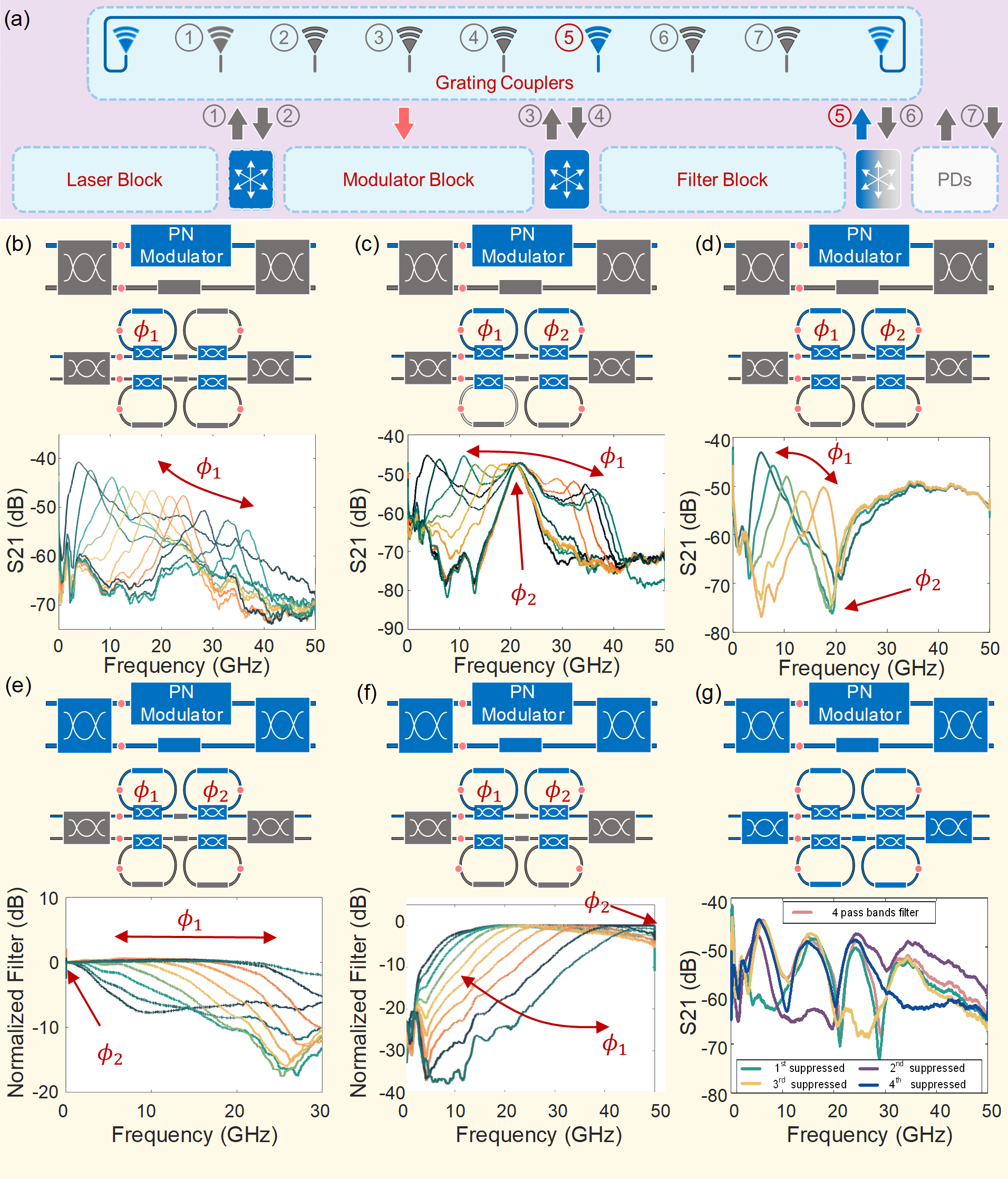}
    \caption{(a) The signal processor configuration for E/E conversion. (b) Microwave filter: single tunable bandpass filter. (c) Microwave filter: double tunable bandpass filter; (d) Microwave filter: tunable one-pass one-stop band filter; (e) Microwave filter: tunable low pass filter; (f) Microwave filter: tunable high pass filter; (g) Microwave filter: four tunable bandpass filters.}
    \label{fig:4}
\end{figure}

\noindent\textbf{Electrical-to-optical (E/O) response.} Combining the laser(s) and the modulator, we can imprint RF signals onto the optical carrier(s), corresponding to a pure electrical-to-optical response. The on-chip tunable lasers consist of a Fabry–Pérot (FP)  laser and a ring laser with intra-cavity wavelength filters. The measurements show \qty{-3}{dBm} optical power coupled to fiber with a tuning range of \qty{60}{nm}. Depending on the design of the transfer-printed SOA, the lasing range can be adjusted \cite{Soltanian2022Micro-transfer-printedRange} (details in supplementary information S2.1). The reconfigurable modulator block consists of a Mach-Zehnder modulator with tunable couplers and a PN junction carrier depletion modulator in one arm, and a slow thermal phase shifter in the other arm. With the tunable couplers and phase shifter we can configure the modulator to provide either  intensity modulation or pure phase modulation, which is measured with more than \qty{20}{dB} extinction with the same optical powers in an off-chip PD \cite{Deng2019PureModulator} (details in supplementary information S2.3). 
\\
\noindent\textbf{Optical-to-Electrical (O/E) response.} The O/E conversion is implemented by two on-chip PDs. Because the packaged connectors ofthe packaged chip still introduce excessive RF crosstalk, we introduce an extra erbium-doped fiber amplifier (EDFA) in the signal path (going off-chip through the grating couplers). This boosts the modulated light signal, in turn enhancing the electrical signal. The O/E response of the on-chip PDs shows a \qty{3}{dB} bandwidth of \qty{25}{GHz}. More details are shown in the supplementary information S2.5.
\\
\noindent\textbf{Electrical-to-electrical (E/E) response.} The packaged chip can act as an analog microwave processor and implement sophisticated linear filtering of RF signals, as shown in Fig.~\ref{fig:4}. The response is measured with an off-chip PD to avoid the RF crosstalk altogether. Unlike previously demonstrated integrated microwave photonic filters  using single-sideband modulation (SSB)\cite{Fandino2017AFilter, Guo2021VersatileShaper}, our signal processor can use both modulated sidebands, which simplifies the optical filter design. The reconfigurable modulator can generate sidebands with tunable phase relations (details in supplementary information S2.3), and with the reconfigurable optical filters, the RF response coming from the PDs can be fully controlled. Furthermore, to construct a generalized bandpass or bandstop filter in the RF domain requires only cascaded ring filters. More details explaining the unique benefits of this double-sideband approach can be found in supplementary information S3.1. Figure~\ref{fig:4}b-g shows various types of RF filter responses, realized by tuning both the modulator block and the rings. Limited by the $Q$ factors of the rings, the narrowest \qty{3}{dB} RF filter bandwidth is $\sim$\qty{3}{GHz}.
\\

\begin{figure}[!t]
    \centering
    \includegraphics[width=\linewidth]{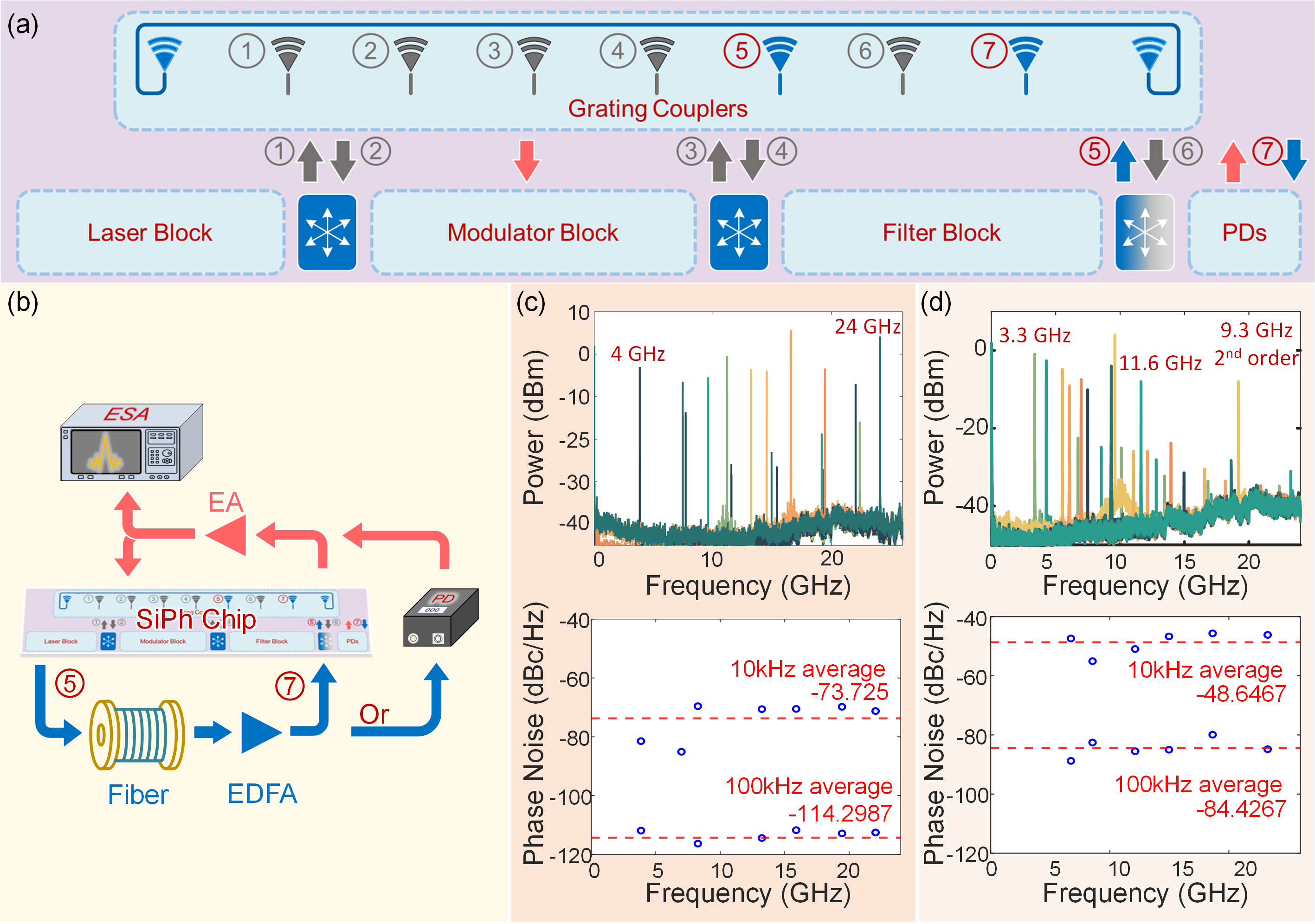}
    \caption{(a) The signal processor configured as an opto-electronic oscillator. (b) Measurement setup for the OEO. (c) RF signal generation with \qty{500}{m} fiber and off-chip PD; (d) RF signal generation with \qty{20}{m} fiber and on-chip PD; ESA: Electrical Spectrum Analyzer. EA: Electronic Amplifier. EDFA: Erbium-Doped Fiber Amplifier.}
    \label{fig:6}
\end{figure}

\noindent\textbf{RF frequency multiplier.} The signal processor reported can implement RF frequency doubling using the laser, the modulator and the detector. By biasing the Mach-Zehnder modulator at its null point\cite{Yao2009MicrowavePhotonics}, the ratio of the second-harmonic signal ($2f$) to the linear response ($1f$) can be adjusted by tuning the couplers in the modulator block such that the losses in both arms are fully balanced (this is a 70:30 balance as the PN modulator has a higher loss). This results in a $2f/1f$ signal ratio that is as high as \qty{40}{dB} with an off-chip PD. For an on-chip PD, the RF crosstalk degrades the $1f$ extinction (details in supplementary information S3.2).
\\

\noindent\textbf{Opto-electronic oscillator (OEO).} With additional electronic amplifiers compensating the loss of the RF bandpass filters shown above, our signal processor can form an OEO with a tunable central frequency, as shown in Fig.~\ref{fig:6}. An extra length of fiber is added in the light path to form a longer oscillation cavity, which is indicated in Fig.~\ref{fig:6}. With an off-chip PD, RF signals can be generated with afrequency tunable from \qty{4}{GHz} to \qty{24}{GHz}, and \qty{-116}{dBc/Hz} phase noise at \qty{100}{kHz} offset frequency, shown in Fig.~\ref{fig:6}c. And with the on-chip PD, the generated signal can be tuned from \qty{3.3}{GHz} to \qty{11.6}{GHz} with \qty{-84}{dBc/Hz} phase noise at \qty{100}{kHz} offset frequency, shown in Fig.~\ref{fig:6}d. More details are shown in supplementary information S3.3.
\\

Except for the functionalities shown above, our signal processor can also work as an optical wavelength meter or RF frequency meter using the tap monitors in the filter block, where the resolution is limited by the $Q$ factor of the loaded ring (Supplementary information S3.4).

\section{Discussion}\label{sec4}

The single-chip signal processor provides a fully programmable filtering response for optical and RF signals, as well as generation and detection of electrical and optical signals. As a demonstrator, the performance of this processor is limited by its system scale, and the packaging and crosstalk management leave room for improvement. 
\\

\noindent\textbf{Microwave photonic filtering based on double-sideband modulation.} The E/E response of the reported signal processor is based on double sideband modulation. Compared to SSB modulation schemes, our algorithm allow a simpler filter design and a lower intrinsic RF loss. The scheme relies on the reconfigurable modulator, which can generate RF sidebands with 
a tunable phase difference, while the tunable couplers ensure the minimum optical loss in all scenarios. In this case, the optical filters used to filter the modulation introduced sidebands can just be as simple as cascaded tunable ring resonators. 
\\

\noindent\textbf{Filtering performance and scale-up analysis.} The optical and electrical filters are both based on the same optical filter block in our signal processor. As a tunable filter, the minimum \qty{3}{dB} bandwidth depends on the quality ($Q$) factor of the rings, which were $\sim 50000$ at critical coupling. The monitor PDs inside the ring introduce extra loss (limiting the $Q$), but they monitor the ring status for configuration. Better calibration schemes or a tunable power tap could allow us to reduce the impact of these monitor PDs. 

The filtering performance is also limited by the filter order, which corresponds to the number of ring filters in the block. For a higher-order RF filter, we only need to cascade additional ring filters, which is very scalable in terms of optical losses and tuning scheme. For optical filtering, increasing the order requires extra ring pairs over both arms of the MZI. 
\\

\noindent\textbf{Equalized E/O conversion with widely tunable optical carriers.} The two on-chip laser sources both show \qty{50}{nm} tuning ranges in a different wavelength region, and a total \qty{90}{nm} tuning range (from \qty{1508}{nm} to \qty{1598}{nm}). With the on-chip modulators, an E/O conversion with widely tunable optical carriers can be reached. 

The programmable filter can also be used as an RF equalizer. In the supplementary information S3.2 we show how we compensate the uneven RF loss of the PCB and bonding wires, boosting the \qty{3}{dB} bandwidth from \qty{4.5}{GHz} to \qty{26}{GHz}.

In addition, the on-chip laser together with two PDs can also be configured as a coherent receiver with tunable local oscillator (LO). But due to the high RF crosstalk (discussed in the crosstalk section), we did not further evaluated this.
\\

\noindent\textbf{RF signal generation.} We reported two demonstrations for RF signal generation. We showed frequency doubling with \qty{40}{dB} $2f/1f$ extinction ratio using the configurable modulator, and a tunable OE oscillator with an external fibre and RF amplifier.  The OEO's RF generation covers the 4-\qty{24}{GHz} span with almost constant phase noise of \qty{-114}{dBc/Hz} at \qty{100}{kHz} using an off-chip PD. This is superior to commercial products (R\&S SMR, \qty{-105}{dBc/Hz} at \qty{100}{kHz} offset frequency for \qty{10}{GHz} central frequency). The frequency generation of the OEO system is not limited to \qty{24}{GHz} with our signal processor, but its phase noise measurement is limited by the used spectrum analyzer (\qty{26}{GHz}). 
\\

\noindent\textbf{Thermal crosstalk.} For configurability, the  system contains 52 thermal tuners, making it  quite sensitive to external and internal temperature fluctuations. In addition, the power dissipation of the transfer-printed laser (\qty{300}{mW}) and the modulator termination (\qty{80}{mW} at \qty{-2}{V}) make the thermal control even more challenging. Even though the sample is mounted on a temperature controller, it still drifts with ambient temperature. Apart from this, the system shows a high stability and repeatability, which means that the different crosstalk contributions can be compensated with the tuners.
\\

\noindent\textbf{Optical crosstalk.}
The chip has an integrated laser, but no optical isolator. This makes the laser cavities vulnerable to all  reflections along the optical path. As these are not fully avoidable, all phase tuners will also affect the reflected light, introducing small perturbations to the laser cavity. However, the tunability of the system allows us to compensate for most of these. \\

\noindent\textbf{RF crosstalk.}
A key limitation to our demonstration reported here is the RF crosstalk. This can be mostly attributed to the packaging: the high-speed PCB couples the RF signal  directly from the input connectors to the output connectors, bypassing the chip. Because we did not boost the PD output with transimpedance amplifiers (TIAs), their relatively weak signal is swamped by  the direct RF crosstalk and cannot be retrieved. For the reported experiments we always used an off-chip EDFA to boost the modulated optical signal and then fed it back into the chip. With a proper TIA and better RF isolation, the extra EDFA would not be necessary. Details of this characterization are shown in the supplementary information S2.5.

\section{Conclusion}\label{sec5}

In this work, we reported the first single-chip microwave photonic signal processor based on silicon photonics. It can be programmed as a reconfigurable filter for optical signals and RF signals, and provide E/O and O/E conversions. The laser sources based on transfer printing make it work without an external light source. Using multiple monitor detectors, the system can be configured locally. In addition, the processor shows its capabilities for RF signal generation and RF frequency doubling. In general, this demonstration suggests a promising and comprehensive approach for the generation, distribution, and processing of optical and RF signals and its potential use in data centers, wireless and satellite communication, and other optical and microwave applications.

\backmatter

\bmhead{Supplementary information}

See the attachment file.

\bmhead{Acknowledgments}

Acknowledgments are not compulsory. Where included they should be brief. Grant or contribution numbers may be acknowledged.

The research presented here was supported by the European Research Council through the Consolidator Grant PhotonicSWARM (grant 725555), and the European Horizon2020 program through the projects MORPHIC (grant 780283), CALADAN (grant 825453) and  MedPhab (grant 871345). 

\bmhead{Author Contributions} H.D. and W.B. conceived the idea of the project. H.D. and J.Z. simulated and designed the whole system and laid out the mask.  H.D. performed the experimental characterizations and analysis. E.S. helped with the transfer printing process. X.C. helped with the mask layout. C.P. helped with the chip post processing. N.V., D.N., J.R, and J.D provided the SOA coupon fabrication. N.S. and G.T. helped with the OEO characterization. H.D., E.S. and W.B. wrote the paper. G.R and W.B. supervised the project. All authors commented on the manuscript.

\section*{Declarations}
The authors declare no conflicts of interest

\bibliography{references.bib}

\section{Methods}\label{sec3}

\noindent\textbf{Silicon chip Fabrication.} The photonic chip was fabricated in the imec iSiPP50G silicon photonic platform with \qty{200}{mm} wafers. The chip size is $1.3 \times \qty{5}{mm^2}$ without grating couplers and $3.3 \times \qty{5}{mm^2}$ with grating couplers. The chip layout is designed using IPKISS by Luceda Photonics. 
\\

\noindent\textbf{Transfer printing process.} Transfer printing technology is using a stamp to pick up a prefabricated SOA coupon from a III-V source wafer and print it on to the target position of the silicon photonic sample (or wafer). In this process, SOAs are fabricated on a 2" wafer of InP epitaxial layer stack, grown by metal-organic vapor-phase epitaxy (MOVPE) method at III-V Lab. A release layer is incorporated beneath the SOA device layer stack. After devices are patterned, the release layer is selectively etched, resulting in free-standing components held by tethers. A polydimethylsiloxane (PDMS) stamp is then laminated against these released devices, termed "coupons". When the stamp is rapidly pulled away, the coupons adhere due to strong bonding, break the tethers, transferring the devices to the stamp.

After pick-up, the stamp now carrying the device (either single or array), is aligned to the destination wafer using pattern recognition through the transparent PDMS stamp. The printing process entails laminating the stamp to the target wafer and retracting it slowly, using a shear force, ensuring the coupons adhere to the target wafer. To enhance coupon adhesion, an divinylsiloxane bisbenzocyclobutene (DVS-BCB) adhension layer is used. After the process, one metallization step is necessary to link the transfer-printed components with the bondpads on the original silicon photonic sample \cite{Roelkens2023Micro-TransferCircuits}.

On the imec iSiPP50G platform, a localized back-end opening (recess) is necessary to interface with the Si waveguide, permitting the seamless integration of III-V SOAs with Si structures. The 40 µm wide III-V SOA measures 1 mm, including two 180 µm adiabatic tapers to enhance the coupling efficiency. Underneath the SOA, there's a continuous poly-Si/c-Si waveguide (with respective thicknesses of 160 nm and 220 nm) ensuring optimal optical coupling between the III-V device layer and the Si waveguide below it. Moreover, an added taper structure aids in channeling the optical mode to the 220 nm thick crystalline Si wire waveguide. Detailed information regarding the design and production of the III-V/Si SOAs can be found in \cite{Soltanian2022Micro-transfer-printedRange}, and more details can be seen in Supplementary Information S1.2.
\\

\noindent\textbf{Optical, electrical and RF Packaging.} Fiber arrays are used for coupling the light in and out of chip. A fiber array is actively aligned by maximizing transmission through a shunt waveguide  and  then fixed in place on the processed sample using UV-curable epoxy. A \qty{15}{dB} loss is measured for the reference waveguide ports, corresponding to a \qty{7.5}{dB} loss per grating coupler. The use of edge couplers instead of grating couplers could significantly reduce the insertion loss of the fiber interfaces \cite{Ferraro2023ImecRoadmap}. 

The 68 DC channels (heaters, tap monitors and grounds) and 3 RF ports (1 input, 2 outputs) are wirebonded to a high-speed printed circuit board (PCB). More details are shown in the supplementary information S1.3. However, the modulator and the PD performance show that the wirebonding to the transmission line introduces a high loss and high RF crosstalk, which limits the maximum frequency that the full system can support to \qty{25}{GHz}. A better RF packaging solution may relax these constrains and bring the performance up to the level of what we measured using RF probes: \qty{33}{GHz} for the modulator and \qty{50}{GHz} for the PDs.
\\

\noindent\textbf{Electrical driving system.} The transfer-printed SOA, the tap monitors and the on-chip PDs are driven by a Tektronix Keithley 2401. The heater-based optical phase shifters for tuning and configuration are driven by a PXI 6704 from National Instruments. 
\\

\noindent\textbf{Optical response measurement.} No external laser is used to generate the different functions described above. The transmission spectrum for the O/O functionality was measured with an optical network analyzer (LUNA, OVA5000), which does contain a tunable laser, but which is only used for characterisation purposes. The off-chip PD is a \qty{42}{GHz} PD (Discovery LabBuddy DSC10H, \qty{0.54}{A/W} at \qty{1550}{nm}). 
\\

\noindent\textbf{Electrical response measurement.} The E/E response is measured by a Keysight vector network analyzer(E8364B, \qty{50}{GHz}). The RF signal source in Fig.~5(b) is a \qty{40}{GHz} signal generator (Rohde\&Schwarz SMR 40) and the electrical spectrum analyzer in Fig.~5(b) is a Keysight EXA signal analyzer (N9010A \qty{44}{GHz}). The ESA used in Fig.~6(b) for the spectrum and phase noise measurement is an Anritsu MS2840A (\qty{26}{GHz}).



\section*{Supplementary material}

# Single-Chip Silicon Photonic Processor for Analog Optical and Microwave Signals
## Supplementery material

Hong Deng, Jing Zhang, Emadreza Soltanian, Xiangfeng Chen, Chao Pang, Nicolas Vaissiere, Delphine Neel, Joan Ramirez, Jean Decobert, Nishant Singh, Guy Torfs, Gunther Roelkens, Wim Bogaerts

# S1 Chip Fabrication, Post-processing and Packaging

## S1.1 Chip Design and Fabrication

The chip is designed with Luceda IPKISS, and fabricated on the imec iSiPP50G silicon photonic platform. This is a standarized silicon photonics platform based on a 220 nm thick silicon-on-insulator (SOI) layer. It supports three waveguide cross sections (strip, rib and deep rib), multiple implantations for high-speed modulators (up to 50 GHz in some configurations), germanium photodetectors (PDs), and two layers of Cu damascene metallization. Our chip design and fabrication were based on the standard process flow which is also offered through multi-project wafer services [4, 7].

## S1.2 Post-processing

The on-chip optical gain is provided by transfer-printed prefabricated semiconductor optical amplifiers (SOAs). III-V SOAs in this work are fabricated based on two epitaxial layer stacks with two multi-quantum wells (MQWs) zone variants to achieve photoluminescence (PL) peaks around 1500 nm and 1550 nm. Using these two epi-stacks in this work leads to laser emissions around 1525 nm and 1575 nm. The micro-transfer-printing (µTP) concept is illustrated in the Fig. S1.1

Prior to the µTP, a combination of dry-etch (by RIE) and wet-etch (by BHF) was first applied to the photonic chip to remove the back-end stack containing 2.5 µm thick silicon dioxide on the Si-waveguide, to form the recess where the InP-based SOA will be integrated. The locally opened recess is slightly longer and wider than the pre-fabricated III-V SOAs. Next, a thin divinylsiloxane bisbenzocyclobutene (DVS-BCB) adhesive layer with a thickness of 150 nm was spray-coated to enhance the bonding

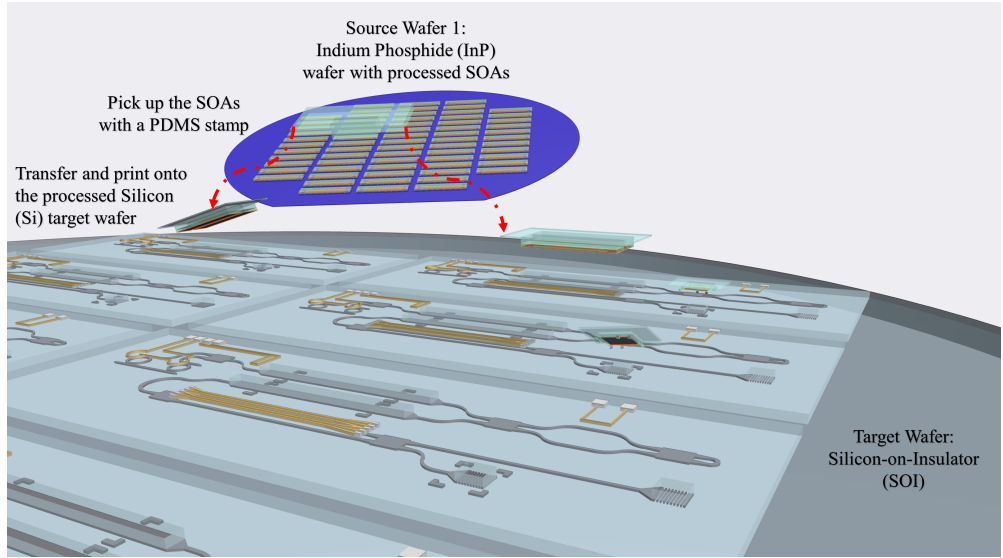

**Fig. S1.1** The general schematic illustration of µTP concept. SOA coupons from source wafer are transfer-printed on desired locations on the target wafer using PDMS stamp.

strength between the III-V SOA and the underlying Si-waveguide. During the spray-coating, DVS-BCB will build up at the edges of the cavity allowing the metallization in the last step to run over the sidewall of the recess, which here is about 4.5 µm high. µTP of III-V SOAs was done using an X-Celeprint µTP-100 lab-scale printer. The post-printing processing finishes by electrically connecting the printed SOAs to the SiPh back-end using 1 µm thick Au on top of a 40 nm Ti layer.

## S1.3 Chip Packaging

The sample is wirebonded to a printed circuit board (PCB) and cured with a fiber array. The photonic integrated circuit features a total of 15 optical input/output ports, 3 RF ports, 52 thermal tuners, 8 monitor PDs, and a transfer-printed laser. As a result, a comprehensive optical, DC, and RF packaging approach is essential to ensure the stability and robustness of the system.

A printed circuit board (PCB) was designed for the electrical connections. The processor chip was placed face-up on the board, and the bondpads were connected using wirebonding.

The optical packaging process involves the attachment of a fiber array. Specifically, a 32-channel fiber array was positioned with active alignment on the reference waveguide, and then fixed in place with UV-curable glue. The transmission of the reference waveguide was measured with an optical vector analyzer (LUNA, OVA 5000), and the setup is presented in Fig. S1.2(a), with the corresponding plot shown in Fig. S1.2(b).

During the measurement, the OVA swept the wavelength spectrum while considering both polarizations. However, due to the grating coupler's limitation in supporting

only the TE mode, we observe an additional 3 dB loss incurred for the polarization averaging. Consequently, the coupling loss per pair of grating couplers was determined to be 14.7 dB, with the peak transmission at 1543.3 nm.

To encapsulate the DC and RF wires, a significant amount of UV glue was added during the optical packaging process. Unfortunately, this resulted in a somewhat uneven curing process, leading to higher optical losses than normal. Nevertheless, this measured transmission curve will be used as the normalization reference for all optical responses in subsequent sections (Section S2.1 and S2.4).

The high speed PCB for the packaging is a 4-layer stacked PCB with high-speed materials (Megtron 6) as dielectric layers. The used RF connectors are Rosenberger solderless RF connectors. A straight referencing transmission line (2 cm) shows a 3 dB drop over 40 GHz between two connectors, as shown in Fig. S1.3(a). To evaluate the effect of bondwires, we wirebonded a pair of transmission lines, and got the transmission response (half of two transmission line with a bonding wire) as shown in Fig. S1.3(b). The curve in Fig. S1.3(b) is used as reference for the modulator tests (Section S2.3). But because of the different bonding targets, the wirebonding responses for our signal processor cannot be fully calibrated. On the other hand, we observe that the PCB introduces RF crosstalk, which couples the RF signal directly from the inputs (of the modulator) to the outputs (of the PD). This measured transmission response (with an 20 dB amplifier, the actual crosstalk is 20 dB lower) is shown in the Fig. S1.3(c). This curve is later used in the characterization of the on-chip PDs (Section S2.5).

## S2 Chip Functional Block Characterization

Our single-chip signal processor contains a laser block, a reconfigurable modulator block, an optical filter block, and two high-speed PDs. Optical switches are used for connecting the blocks, and guiding the light in/out of the optical path. Their characterizations are shown in this section.

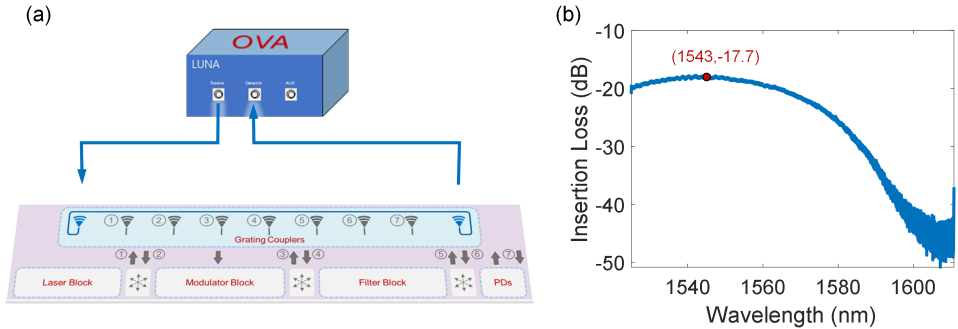

**Fig. S1.2** (a) The measurement setup for the insertion loss of the reference waveguide. (b) The measured results.

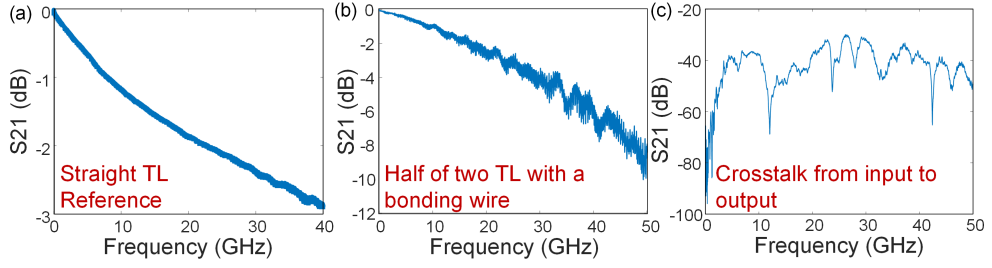

**Fig. S1.3** (a) Transmission of a reference straight transmission line; (b) Half transmission of two used transmission line with a bonding wire; (c) The direct crosstalk from input RF connector to output RF connector.

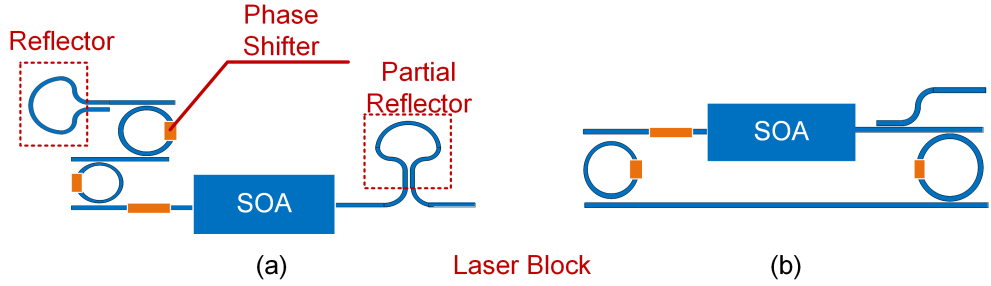

**Fig. S2.1** Schematic of the laser block. (a) A Fabry–Pérot cavity laser; (b) A ring cavity laser. SOA: Semiconductor optical amplifier.

## S2.1 Laser Block: Transfer-Printed Lasers

In our signal processor, there are two different laser cavities designed in the laser block, which are based on a Fabry–Pérot (FP) cavity and a ring cavity laser, respectively, as shown in Fig. S2.1. Two ring resonator filters are used in each laser cavity to form a Vernier bandpass filter to select the longitudinal modes. Prefabricated SOAs are then transfer printed into the laser cavities to provide optical gain. When the gain from the printed SOA is higher than the other accumulated losses in the laser cavity, lasing will start.

There are three thermal phase shifters in each laser cavity: two for the phase tuning in the ring and one for the cavity phase. By tuning the phase of the rings and optical cavity, we can tune the lasing wavelength. The laser tuning results are shown in Fig. S2.2. To be noted is that the FP laser of the packaged sample turned out to be electrically shorted after the wirebonding, so the results shown here had to be obtained from another sample. The two SOAs printed in these two laser cavities have different photoluminescence (PL) peaks around 1500 nm and 1550 nm, leading to lasing peaks around 1525 nm and 1575 nm, respectively. From Fig. S2.2, both laser cavities can achieve a tunability of more than 50 nm, and the full tuning range of the laser block can reach up to 90 nm.

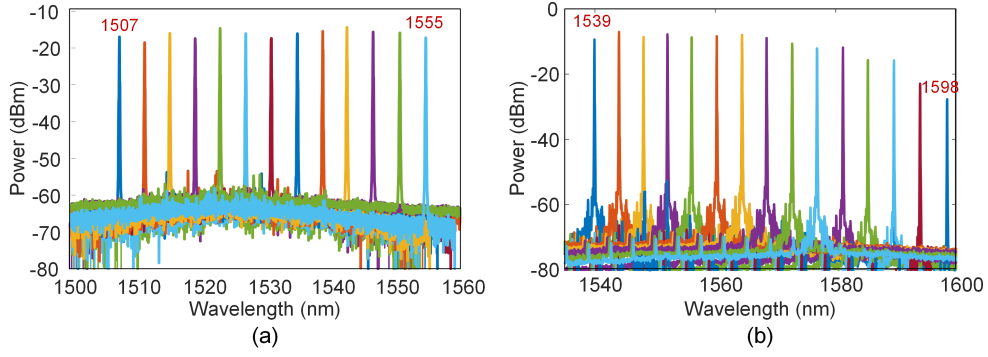

**Fig. S2.2** Laser tuning results. (a) Using the Fabry–Pérot cavity laser (With another sample); (b)Using the ring cavity laser. The results are measured from grating couplers.

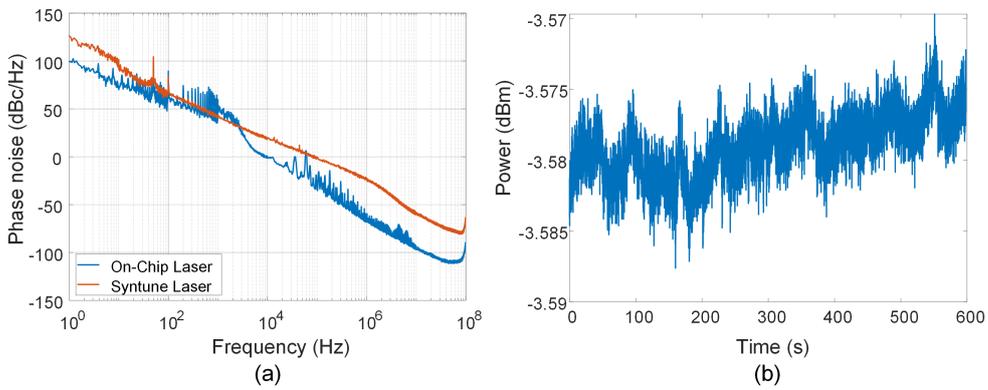

**Fig. S2.3** (a) Phase noise of the ring laser. (b) Stability test of the ring laser.

The phase noise of the ring laser is shown in Fig. S2.3(a), which is measured by an OEwave OE4000 optical phase noise test system. The estimated instantaneous linewidth is 283 kHz. The orange curve is a commercial laser source (Syntune laser). The on-chip laser shows a overall lower phase noise. The stability of the output power is shown in Fig. S2.3(b), from which we can find that the output power of the laser varies with only 0.015 dB within a 600 second timeframe.

### S2.2 Optical Switches and Tunable Couplers

Optical switches are used to connect the functional blocks on the chip and guide light into / out of the optical path. In addition, tunable couplers are used in the modulator block and optical filter block to tune the relative power in each path with specific coupling coefficients. In our designs, the optical switches and tunable couplers are achieved by single- and double-stage Mach-Zehnder interferometers (MZIs).

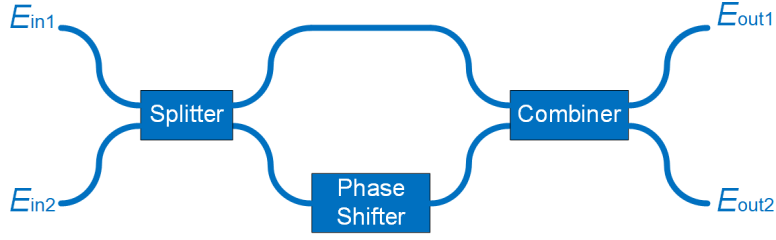

**Fig. S2.4** A single Mach-Zehnder interferometer.

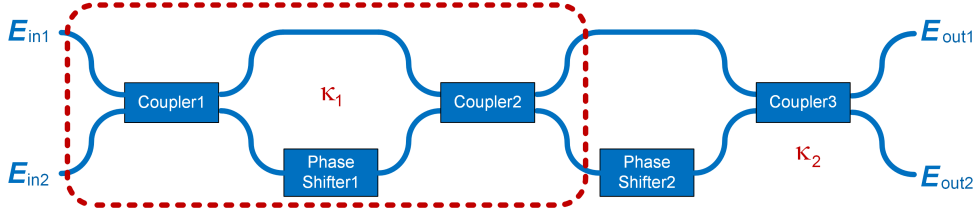

**Fig. S2.5** A double-stage MZI as a tunable coupler.

A schematic for a single stage MZI is shown in Fig. S2.4, which includes an optical power splitter, an optical combiner and one phase shifter. The power splitter and the combiner used in our processor are directional couplers (DCs) and the phase shifter is a thermal phase shifter. Then the optical output field can be expressed as:

$$\begin{bmatrix} E_{\text{out1}} \\ E_{\text{out2}} \end{bmatrix} = \begin{bmatrix} \sqrt{(1-\kappa)} & j\sqrt{\kappa} \\ j\sqrt{\kappa} & \sqrt{(1-\kappa)} \end{bmatrix} \begin{bmatrix} 1 & 0 \\ 0 & e^{-j\phi_s} \end{bmatrix} \begin{bmatrix} \sqrt{(1-\kappa)} & j\sqrt{\kappa} \\ j\sqrt{\kappa} & \sqrt{(1-\kappa)} \end{bmatrix} \begin{bmatrix} E_{\text{in1}} \\ E_{\text{in2}} \end{bmatrix} \quad (S2.1)$$

where $E_{\text{out1}}$ and $E_{\text{out2}}$ are the electrical field of the MZI's outputs, $E_{\text{in1}}$ and $E_{\text{in2}}$ are the inputs' electrical field, $\kappa$ is the splitting ratios of the splitter and combiner, and $\phi_s$ is the phase shift introduced by the phase shifter.

If perfect 50:50 DCs are used for the splitter and combiner, the transfer function from $E_{\text{in1}}$ to $E_{\text{out1}}$ and from $E_{\text{in1}}$ to $E_{\text{out2}}$ can be expressed as:

$$\boldsymbol{r}_{\text{MZI}} = \frac{E_{\text{out1}}}{E_{\text{in1}}} = \frac{1}{2}[1 - e^{-j\phi_s}] = \sin(\frac{\phi_s}{2})e^{-j\frac{\phi_s}{2}} \quad (S2.2)$$

$$\boldsymbol{\kappa}_{\text{MZI}} = \frac{E_{\text{out2}}}{E_{\text{in1}}} = \frac{1}{2}j[1 + e^{-j\phi_s}] = \cos(\frac{\phi_s}{2})e^{-j(\frac{\phi_s}{2} - \frac{\pi}{2})} \quad (S2.3)$$

From here we can notice that the output signal's magnitude follows a sinusoidal relation of the half of the $\phi_s$, and the output signal's phase is shifted also by half of the $\phi_s$. And we can also get the power transformation of the MZI, which is:

$$|\boldsymbol{r}_{\text{MZI}}|^2 = \frac{1 - \cos(\phi_s)}{2} \quad (S2.4)$$

$$|\boldsymbol{\kappa}_{\text{MZI}}|^2 = \frac{1 + \cos(\phi_s)}{2} \quad (S2.5)$$

It can be noted that the power coupling ratios $|\boldsymbol{r}_{\text{MZI}}|^2$ and $|\boldsymbol{\kappa}_{\text{MZI}}|^2$ fully depend on the phase shifts $\phi_s$, which makes it an *optical tunable coupler* or *optical switch*[6]. When $\phi_s = 0$, the

two arms of the MZI are fully balanced, and all light from input port 1 is coupled to output port 2, which is called *cross* state, and when the phase difference $\phi_s = \pi$, the light from input 1 is fully coupled to output 1, called *bar* state.

If the DCs do not have a perfect 50:50 splitting ratio due to the fabrication errors or wavelength dependence, the overall power tuning of the MZI, $|\boldsymbol{r}_{\mathrm{MZI}}|^2$ and $|\boldsymbol{\kappa}_{\mathrm{MZI}}|^2$, cannot be fully tuned of the range of $0 - 1$ with the phase tuning $\phi_s$. In that case, a double-staged MZI can be used to release the constraints [8] . With the scheme shown in Fig. S2.5, the Eq. S2.1 can be rewritten as:

$$\begin{bmatrix} E_{\mathrm{out1}} \\ E_{\mathrm{out2}} \end{bmatrix} = \begin{bmatrix} \sqrt{(1-\kappa_2)} & j\sqrt{\kappa_2} \\ j\sqrt{\kappa_2} & \sqrt{(1-\kappa_2)} \end{bmatrix} \begin{bmatrix} 1 & 0 \\ 0 & e^{-j\phi_2} \end{bmatrix} \begin{bmatrix} \sqrt{(1-\kappa_1)} & j\sqrt{\kappa_1} \\ j\sqrt{\kappa_1} & \sqrt{(1-\kappa_1)} \end{bmatrix} \begin{bmatrix} E_{\mathrm{in1}} \\ E_{\mathrm{in2}} \end{bmatrix} \quad (S2.6)$$

where $\kappa_2$ is the power coupling ratio of the normal coupler, $\kappa_1$ is the power coupling ratio of the tunable coupler, and $\phi_2$ is the phase shift from the shifter 2 in figure S2.5. Then the field transfer function can be written as:

$$\boldsymbol{r}_{\mathrm{MZI}} = \sqrt{(1-\kappa_1)(1-\kappa_2)} - \sqrt{\kappa_1\kappa_2}e^{-j\phi_2} \quad (S2.7)$$

$$\boldsymbol{\kappa}_{\mathrm{MZI}} = j\sqrt{(1-\kappa_2)\kappa_1} + j\sqrt{\kappa_2(1-\kappa_1)}e^{-j\phi_2} \quad (S2.8)$$

Then the optical power transfer functions become:

$$|\boldsymbol{r}_{\mathrm{MZI}}|^2 = (1-\kappa_1)(1-\kappa_2) - 2\sqrt{\kappa_1\kappa_2(1-\kappa_1)(1-\kappa_2)}\cos(\phi_2) + \kappa_1\kappa_2 \quad (S2.9)$$

$$|\boldsymbol{\kappa}_{\mathrm{MZI}}|^2 = \kappa_1 + \kappa_2 - 2\kappa_1\kappa_2 + 2\sqrt{\kappa_1\kappa_2(1-\kappa_1)(1-\kappa_2)}\cos(\phi_2) \quad (S2.10)$$

Here we can see that $|\boldsymbol{r}_{\mathrm{MZI}}|^2 + |\boldsymbol{\kappa}_{\mathrm{MZI}}|^2 = 1$. In order to get a full tuning range of coupling ratio $[0, 1]$ for the equations S2.9 and S2.10, we can set that $|\boldsymbol{r}_{\mathrm{MZI}}|^2 \in [0, 1]$, thus we get:

$$(1-\kappa_1)(1-\kappa_2) - 2\sqrt{\kappa_1\kappa_2(1-\kappa_1)(1-\kappa_2)} + \kappa_1\kappa_2 = 0 \quad (S2.11)$$

$$(1-\kappa_1)(1-\kappa_2) + 2\sqrt{\kappa_1\kappa_2(1-\kappa_1)(1-\kappa_2)} + \kappa_1\kappa_2 = 1 \quad (S2.12)$$

Then from equation S2.11, we can get:

$$\kappa_1 + \kappa_2 = 1 \quad (S2.13)$$

and from equation S2.12, we can get:

$$\kappa_1 = \kappa_2 \quad (S2.14)$$

Here, we can know that, if $\kappa_1 + \kappa_2 = 1$, this circuit can be tuned to have a high extinction ratio or closer to 0 at bar state ($|\boldsymbol{r}_{\mathrm{MZI}}|^2 = 0$) and if $\kappa_1 = \kappa_2$, this circuit can be tuned to have a zero loss or closer to 1 at bar state ($|\boldsymbol{r}_{\mathrm{MZI}}|^2 = 1$). In actual fabricated devices, there will be some insertion losses for the DCs and waveguides. This will increase the overall insertion loss of the tunable coupler, and slightly drift the working conditions of $\kappa_1$ and $\kappa_2$ for compensating loss imbalance.

The first coupler is a tunable coupler discussed above. If we assume that all basic couplers are similar, from equation S2.10 we can get:

$$\kappa_1 = 2\kappa_2(1-\kappa_2)(1+\cos(\phi_1)) \quad (S2.15)$$

where $\phi_1$ is the phase shift from the shifter 1 in figure S2.5. Substituting equation S2.15 into S2.13 and S2.14 we can get:

$$\frac{1}{2\kappa_2} = 1 + \cos(\phi_1) \quad (S2.16)$$

$$\frac{1}{2(1-\kappa_2)} = 1 + \cos(\phi_1) \quad (S2.17)$$

For a tunable $\phi_1$, we can get:

$$\kappa_2 \in [0.25, 0.75] \quad (S2.18)$$

which means that if the power coupling ratio of the fabricated couplers locates in $[0.25, 0.75]$, the full double-stage MZI's power coupling ratio can be tuned from 0 to 1, by tuning the two phase shifters.

## S2.3 Modulator Block: Reconfigurable modulator

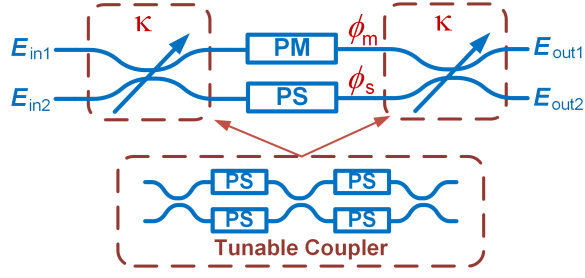

**Fig. S2.6** Scheme of the modulator block. PM: Phase modulator. PS: Phase shifter. TC: Tunable coupler.

The reconfigurable modulator is an MZI formed with two tunable couplers, a PN junction based high speed modulator, two thermal phase shifters and two tap monitors. The scheme is shown in Fig. S2.6. In this modulator design, a full tuning range for the tunable couplers' $\kappa$ is preferred, thus we used double-stage MZI structures to form the tunable coupler. Assuming that the modulator block works in bar state, the modulator's tunable couplers' coupling ratio $\kappa$ are set to be equal to ensure the lowest insertion loss of the whole modulator block ( Eq.S2.14). Then, Eq.S2.1 can be rewritten as:

$$\begin{bmatrix} E_{\text{out1}} \\ E_{\text{out2}} \end{bmatrix} = \begin{bmatrix} \sqrt{(1-\kappa)} & j\sqrt{\kappa} \\ j\sqrt{\kappa} & \sqrt{(1-\kappa)} \end{bmatrix} \begin{bmatrix} \alpha e^{-j\phi_m} & 0 \\ 0 & e^{-j\phi_s} \end{bmatrix} \begin{bmatrix} \sqrt{(1-\kappa)} & j\sqrt{\kappa} \\ j\sqrt{\kappa} & \sqrt{(1-\kappa)} \end{bmatrix} \begin{bmatrix} E_{\text{in1}} \\ E_{\text{in2}} \end{bmatrix} \quad (S2.19)$$

where $\alpha$ is the loss factor of the PN modulator, $\phi_s$ represents the relevant phase difference of the two arm (tuned by the phase shifter), and $\phi_m$ is the modulator introduced phase shift, which can be expressed as:

$$\phi_m = \frac{\pi}{V_\pi} V_{\text{rf}} \cos(\omega_{\text{rf}} t + \phi_{\text{rf}}) \quad (S2.20)$$

In which the $V_{\text{rf}}\cos(\omega_{\text{rf}} t + \phi_{\text{rf}})$ is the modulation RF signal.

Here we assume that the loss of the modulator is unrelated with the modulation signal to simply the modulated signal's expression, which means the modulator is a pure phase modulator. For an actual PN junction based phase modulator, there will be a slightly spurious intensity modulation, which will be discussed later. The equation derivation also holds for actual devices, while all configurations will be slightly different.

Assuming the light signal $E_{\text{in1}} = e^{-j(\omega_c t)}$ is fed from input 1, the modulated light signal at bar port can be expressed as:

$$E_{\text{out1}} = [(1-\kappa)\alpha e^{-j\phi_m} - \kappa e^{-j\phi_s}]E_{\text{in1}} \quad (S2.21)$$

Using the Jacobi–Anger expansion,

$$\begin{aligned} e^{-j\phi_m} &= e^{-j[\pi \frac{V_{\text{rf}}}{V_\pi} \cos(\omega_{\text{rf}} t + \phi_{\text{rf}})]} \\ &= \sum_{n=-\infty}^{\infty} i^n J_n\left(\frac{\pi}{V_\pi} V_{\text{rf}}\right) e^{-jn(\omega_{\text{rf}} t + \phi_{\text{rf}})} \end{aligned} \quad (S2.22)$$

where $J_n$ is is the $n$-th Bessel function of the first kind, we can expand Eq.S2.21 as:

$$E_{\text{out1}} = [(1-\kappa)\alpha \sum_{n=-\infty}^{\infty} j^n J_n(\frac{\pi}{V_\pi}V_{\text{rf}})e^{-jn(\omega_{\text{rf}}t+\phi_{\text{rf}})} - \kappa e^{-j\phi_s}]e^{-j(\omega_c t)}$$

$$\simeq [(1-\kappa)\alpha J_0(\frac{\pi}{V_\pi}V_{\text{rf}}) - \kappa e^{-j\phi_s}]e^{-j(\omega_c t)} + \qquad (S2.23)$$

$$j(1-\kappa)\alpha J_1(\frac{\pi}{V_\pi}V_{\text{rf}})(e^{-j((\omega_c-\omega_{\text{rf}})t-\phi_{\text{rf}})} + e^{-j((\omega_c+\omega_{\text{rf}})t)\phi_{\text{rf}}})$$

$$= A_0 e^{-j\phi_{A_0}} e^{-j(\omega_c t)} + jA_1(e^{-j((\omega_c-\omega_{\text{rf}})t-\phi_{\text{rf}})} + e^{-j((\omega_c+\omega_{\text{rf}})t+\phi_{\text{rf}})})$$

Here only the first sidelobes are kept. The gain of the carrier and left and right modulated sideband are give by

$$A_0 e^{-j\phi_{A_0}} = (1-\kappa)\alpha J_0(\frac{\pi}{V_\pi}V_{\text{rf}}) - \kappa e^{-j\phi_s}$$

$$A_1 = (1-\kappa)\alpha J_1(\frac{\pi}{V_\pi}V_{\text{rf}}) \qquad (S2.24)$$

When the modulated light signal is received by a PD, we can get the photocurrent with the square law detection, which can be expressed as:

$$\begin{aligned}I_{\text{PD}} &= \mathcal{R}E_{\text{out1}}E_{\text{out1}}^* \\ &= \mathcal{R}[A_0 e^{-j\phi_{A_0}} e^{-j(\omega_c t)} + jA_1(e^{-j((\omega_c-\omega_{\text{rf}})t-\phi_{\text{rf}})} + e^{-j((\omega_c+\omega_{\text{rf}})t+\phi_{\text{rf}})})] \\ &\times [A_0 e^{j\phi_{A_0}} e^{j(\omega_c t)} - jA_1(e^{j((\omega_c-\omega_{\text{rf}})t-\phi_{\text{rf}})} + e^{j((\omega_c+\omega_{\text{rf}})t+\phi_{\text{rf}})})] \\ &= \mathcal{R}(A_0^2 + 2A_1^2 - 4A_0 A_1 \sin(A_0)\cos(\omega_{\text{rf}}t + \phi_{\text{rf}}) + 2A_1^2 \cos(2\omega_{\text{rf}}t + 2\phi_{\text{rf}}))\end{aligned} \qquad (S2.25)$$

where $\mathcal{R}$ is the responsibility of the used PD.

Then we can know that the field amplitude of the recovered RF signal at the original frequency is (with Eq.S2.24):

$$\begin{aligned}I_{1^{st}} &= -4\mathcal{R}A_0 A_1 \sin(\phi_{A_0})\cos(\omega_{\text{rf}}t + \phi_{\text{rf}}) \\ &= 4\mathcal{R}\kappa(1-\kappa)\alpha J_1(\frac{\pi}{V_\pi}V_{\text{rf}})\sin(\phi_s)\cos(\omega_{\text{rf}}t + \phi_{\text{rf}})\end{aligned} \qquad (S2.26)$$

which is determined by the coupling ratio of the tunable couplers $\kappa$ and the phase shifter's phase change $\phi_s$.

The recovered RF signal at the original frequency can also be expressed as:

$$\begin{aligned}I_{1^{st}} &= -4\mathcal{R}A_0 A_1 \sin(A_0)\cos(\omega_{\text{rf}}t + \phi_{\text{rf}}) \\ &= -2\mathcal{R}A_0 A_1 [\sin(\omega_{\text{rf}}t + \phi_{\text{rf}} + \phi_{A_0}) + \sin(-\omega_{\text{rf}}t - \phi_{\text{rf}} + \phi_{A_0})] \\ &= -2\mathcal{R}A_0 A_1 [\sin(\omega_{\text{rf}}t + \phi_{\text{rf}} + \phi_{A_0}) - \sin(\omega_{\text{rf}}t + \phi_{\text{rf}} - \phi_{A_0})]\end{aligned} \qquad (S2.27)$$

This signal consists of the two modulated optical sidebands that are down converted with the optical carrier after the square law detection by the photo diode. The relative phase of the two signals ($2\phi_{A_0}$) can be turned by the phase shifter ($\phi_s$), which can be calculated with Eq. S2.24.

Indeed, when $\kappa = 0$, all light goes through the PN modulator and $\phi_{A_0}$ is zero, no signal will be detected after the photodiode as the light was purely phase modulated.

However, when $\kappa = 0.5$, half of the light goes through the phase modulator and the modulator block is configured as an MZ intensity modulator, which can be verified with Eq. S2.26 when $\kappa(1-\kappa)$ and $\sin\phi_s$ are maximized. To enable both working conditions, full $\kappa$ tuning range is required, such that double-stage MZI structures are employed as tunable couplers.

The modulator block is measured with the setup shown in Fig. S2.7. We used the on-chip laser block and the modulator block to load the RF signal generated by a VNA (Keysight

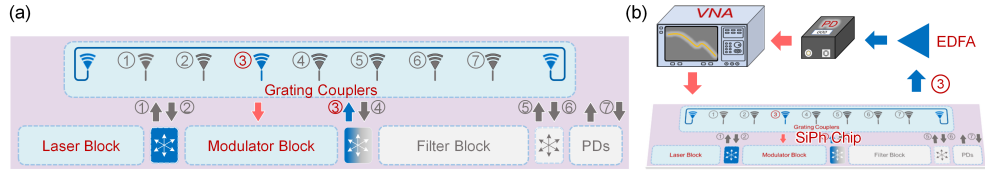

**Fig. S2.7** (a)Sample configuration to test the modulator block. (b) The setup for the measurement. VNA: Vector Network Analyzer; PD: Photodetector. SiPh sample: Silicon photonic sample. EDFA: Erbium-doped fiber amplifier.

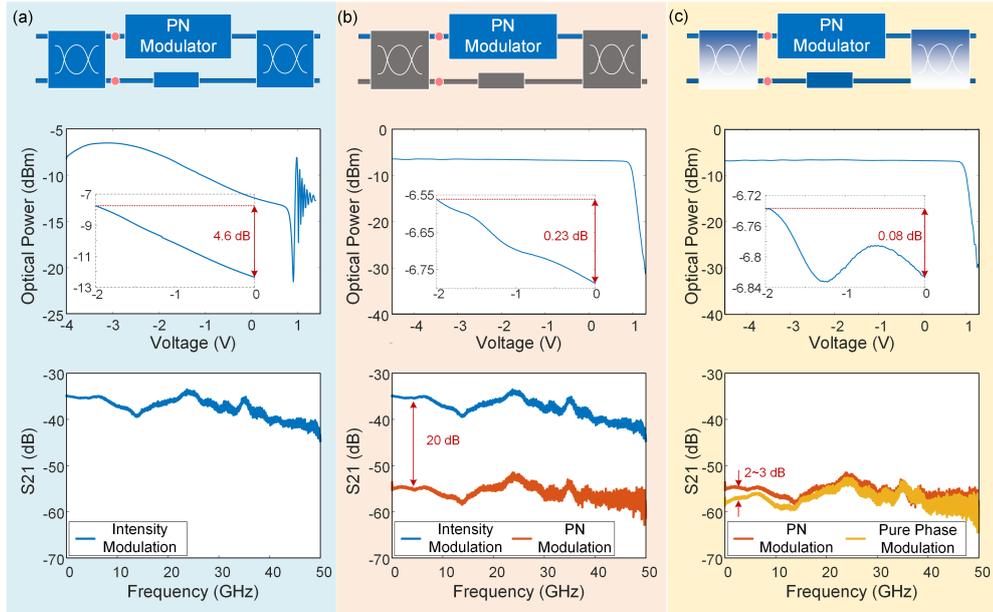

**Fig. S2.8** (a)A configured intensity modulator and its responses; (b) Only the PN junction modulator used, and its spurious intensity modulated response compared to the modulator configured as intensity modulator; (c) A configured pure phase modulator and its reduced spurious intensity modulation. The first row: the configuration of the modulator block. The second row: DC sweep on the PN junction modulator. The third row: the frequency response of the modulator block.

E8364B, 50 GHz) onto an light carrier, and the light signal was coupled out of the chip and received by a standalone off-chip PD (Discovery LabBuddy DSC10H, 43 GHz) to be converted back to a RF signal. The VNA received the recovered RF signal and compare it with what it generated, which corresponds to the S21 scatter parameter of the system under test.

The measured S21 results of the modulator block are shown in Fig. S2.8. The modulator block can be configured as an intensity modulator, a PN junction modulator, and an optimized pure phase modulator. If the tunable couplers' $\kappa = 0$, the modulator block will be simplified to a PN junction based phase modulator. The DC sweep on the PN junction is shown in the middle of Fig. S2.8(b). As can be seen, the PN junction has an absorption variance with the applied voltage at its depletion mode, which will introduce spurious intensity

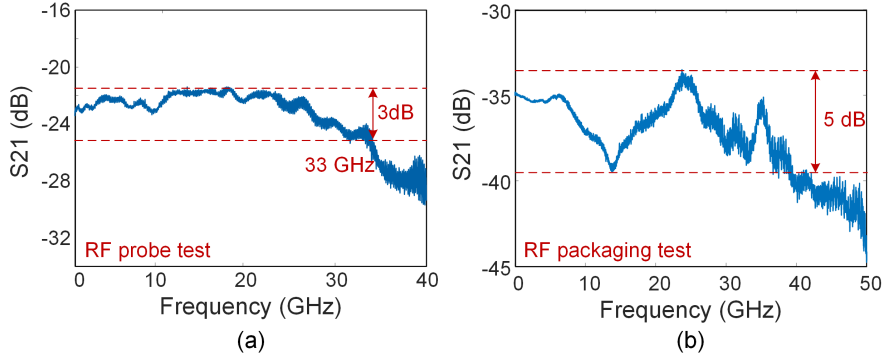

**Fig. S2.9** Modulator frequency response. (a) Tested with RF probes. (b) Tested with full packaging.

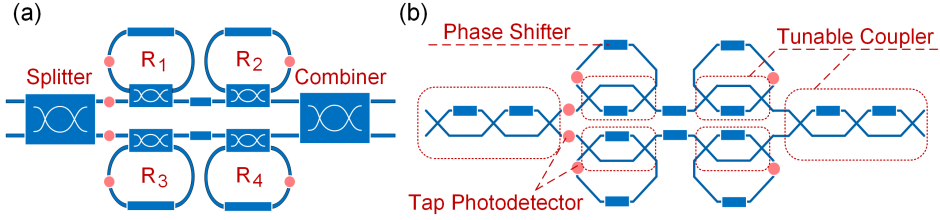

**Fig. S2.10** Schematic of the filter block: Four ring loaded MZI.

modulation. By guiding a little bit light to the other arm of the MZI, this spurious intensity modulation can be suppressed [2], and the response is shown in Fig. S2.8(c). If the tunable couplers' $\kappa$ is set to around 0.5, the modulator block would work as an Mach-Zehnder Modulator for intensity modulation. The responses are shown in Fig. S2.8(a). In this measurement, the output power of the Erbium-doped fiber amplifier (EDFA) is fixed at 7.8 dBm and the photo current of the PD is fixed at 3.2 mA.

The modulator blocks in other samples were also measured with RF probes individually, and they showed good uniformity with a 3 dB bandwidth of 33 GHz as intensity modulators, which matched the data in PDK handbook (33.8 GHz biased at $-1$ V), as shown in Fig. S2.9(a). Compared with the probed modulators, the wirebonding packaged one shows some extra resonance in its response after the calibration, as shown in Fig. S2.8(a) and Fig. S2.9(b). The absolute RF loss difference is due to the optical power difference (10 dBm off-chip laser vs on-chip laser).

## S2.4 Filter Block: Ring-loaded MZI

A multistage ring-loaded MZI with tunable couplers can reach arbitrary magnitude responses by tuning all phase and coupling ratios. In our signal processor, we used a MZI structure with two cascaded rings on both two arms, see in Fig. S2.10. The cascaded rings on each arm work as a second-order all-pass filter (if lossless), and the combiners mixes them up (Eq. S2.6). Classical bandpass filters can be achieved by the sum or difference of two all-pass filters [9]. If the desired response fluctuates between two levels, the bandpass filter design approach can be used with specific MZI coupling ratios, where maximum and minimum of the main MZI spectrum depends on the coupling ratios of the splitter and the combiner (Eq. S2.13

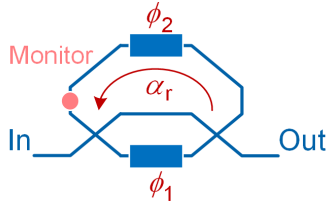

**Fig. S2.11** A ring resonator with a tunable coupler

and Eq. S2.14). A more generic filter design can be achieved with the characteristic function approach [5].

A tunable coupler's transmission can be found as Eq. S2.2 and Eq. S2.3, and the waveguide transmission can be expressed as $\alpha_r e^{-j\phi_2}$ where $\alpha_r$ is the ring's propagation loss factor and the $\phi_2$ is the ring's phase shifts. Then, the transfer function of a ring resonator with a tunable coupler, as shown in Fig. S2.11, can be expressed as:

$$\frac{E_{\text{Out}}}{E_{\text{In}}} = \frac{\sin(\phi_1/2)e^{-j(\phi_1/2+\pi/2)} - \alpha_r e^{-j(\phi_1+\phi_2)}}{1 + \alpha_r e^{-j(\phi_2)}\sin(\phi_1/2)e^{-j(\phi_1/2+\pi/2)}} \quad (S2.28)$$

Because $\alpha_r$ cannot be one (lossless), the ring resonator will not be a perfect all-pass filter and will have a free spectral range (FSR), which can be calculated as [1]:

$$\text{FSR} = \frac{\lambda^2}{n_g L} \quad (S2.29)$$

where $\lambda$ is the optical carrier wavelength, $n_g$ is the group index of the waveguide and $L$ is the round trip length of the ring resonator.

The MZI structure in our signal processor is designed to be balanced, leading to a near-infinite FSR. The ring resonators we used in our signal processor are designed identically, thus their FSR are very similar, and this defines the FSR of the entire filter block. By tuning the splitter and the combiner's coupling ratio $\kappa$ to either 0 or 1, we can measure the transmission of a single ring. The measurement setup is shown in Fig. S2.12(a), and the results are shown in Fig. S2.12(b)-(e). As can be seen, these four rings show an FSR of 0.86 nm or 107.5 GHz. Due to the embedded tunable coupler, the ring's length is relatively large and the FSR is small. A larger FSR can be achieved with a more compact tunable coupler.

The full width at half-maximum (FWHM) or 3 dB bandwidth of the measured ring resonators in their critical coupling states are 30.7 pm, 35.8 pm, 35.8 pm, 43.6 pm, corresponding to quality ($Q$) factors of 50489, 43296, 43296, 35550, respectively. The tap PDs introduce extra loss in the ring cavities, which will limit the Q factor of the rings. The PDs help to monitor the ring status, but they are not strictly necessary if the circuit is fully calibrated and isolated.

The tap PDs couple a fraction of light ($\sim 2\%$) out of the ring and are monitored with an electrical current meter. The monitored photocurrents can help to calibrate the phase shifters ($\phi_1$ and $\phi_2$). We swept the phase shifters' electrical power in the tunable coupler ($\phi_1$) and the ring ($\phi_2$). The PD readings are shown in the Fig. S2.13. As shown in the figure, the peak points indicate that the rings are at their critical coupling states, which can hold the highest optical energy. Then the region in between is the undercoupled state, and the central lowest point is the 0-coupling state, where no light is coupled into the rings. The outer regions are overcoupled states. From these measurements we can also extract that the phase shifter's thermal efficiency is around 28 mW/$\pi$, then we can use this information to fit all phase delays.

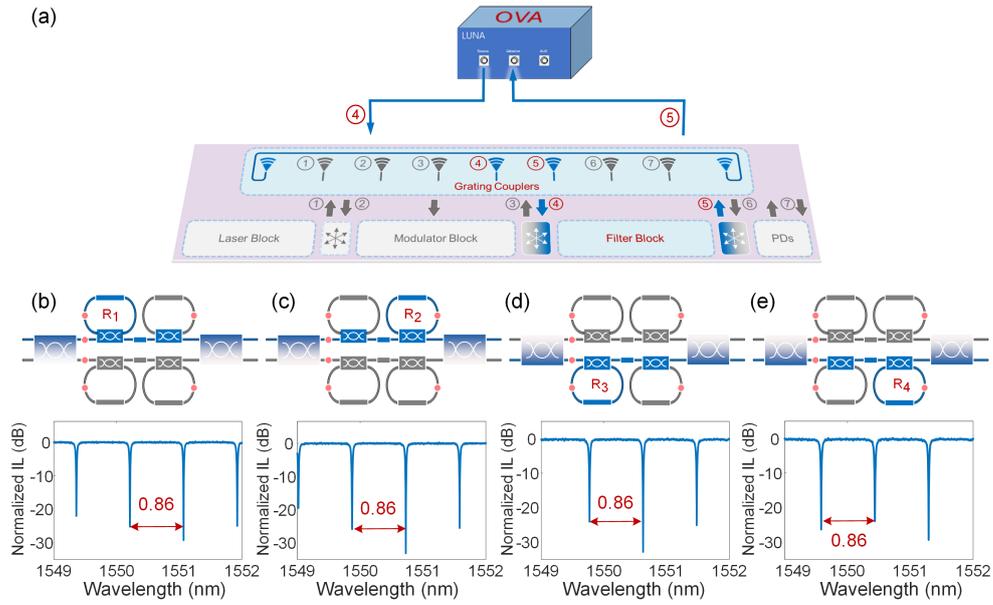

**Fig. S2.12** (a) The measurement setup for the filter block; (b) Ring 1 measurement configuration and results; (c) Ring 2 measurement configuration and results; (d) Ring 3 measurement configuration and results; (e) Ring 4 measurement configuration and results.

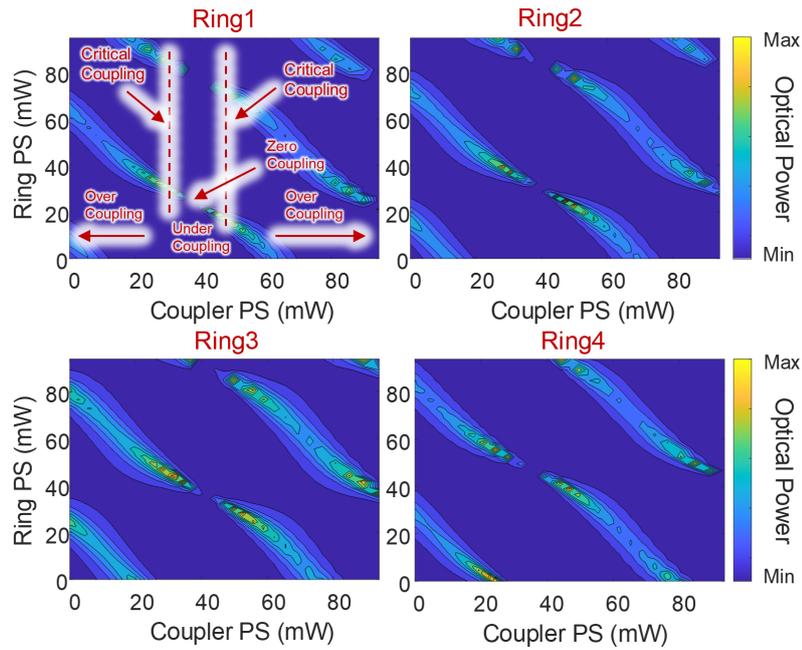

**Fig. S2.13** Tap monitor reading with the phase shifters tuning. Coupler PS: phase shifter in the tunable coupler ($\phi_1$). Ring PS: phase shifter in the ring ($\phi_2$).

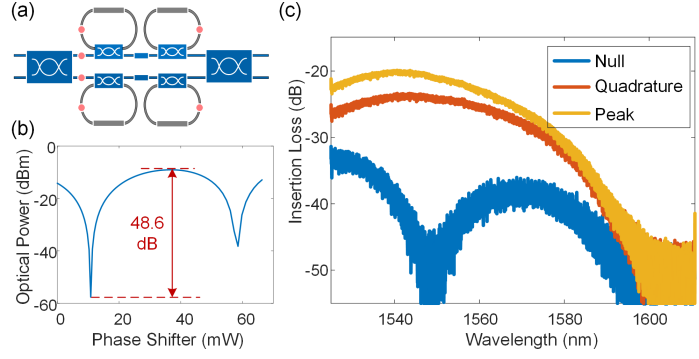

**Fig. S2.14** (a) The configuration of the MZI measurement. (b) The transmission while sweeping the phase shifter on one arm of the MZI. (c) The spectrum response of the MZI biasing at null point, quadrature point and peak point.

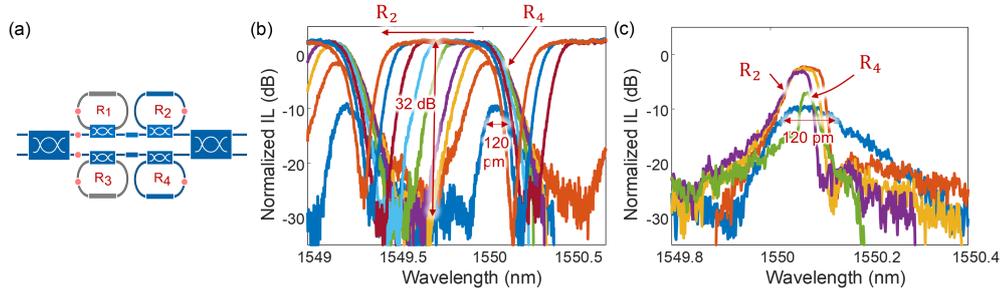

**Fig. S2.15** (a) The configuration of the filter block for a tunable bandpass filter. (b) A tunable bandpass filter response. Two rings are over coupled. (c) A tunable bandpass filter response with narrower passbands. Two rings are tuned from over coupled to under coupled.

With the locally calibrated phase shifters, we can tune the coupling ratios of all rings to 0 to calibrate the main MZI response, which is shown in Fig. S2.14. When sweeping the phase shifter on one MZI arm, the MZI can reach an extinction ratio of 48.6 dB at 1550 nm at cross state, as shown in Fig. S2.14(b). And the optical spectrum responses with different phase states are shown in Fig. S2.14(c). As can be seen, the highest extinction ratio can only be kept in a small wavelength range due to the dispersion of the used directional couplers.

Based on this calibration data, the whole filter block can now be configured with these phase shifters. First, a Chebyshev type II type bandpass filter is defined, as shown in Fig. S2.15. In this filter configuration, only two rings (R2 and R4) are overcoupled to act as all-pass filters, and the other two rings are switched off ($\kappa = 0$), as shown in Fig. S2.15(a). By tuning R2's ring phase, we can get a bandpass filter with a tunable bandwidth, as shown in Fig. S2.15(b). The filter transmission profile shows a passband ripple of 0.5 dB, a roll-off transition of around 210 pm and an extinction ratio of 32 dB. Because of the gradual roll-off, this bandpass filter introduces extra insertion loss when it is too narrow. When the filter bandwidth is narrowed to around 120 pm, the filter introduces 10 dB extra loss. Setting the ring to critical coupling can help to get a sharper roll-off. As can be seen in Fig. S2.15(c), if R2 is kept over coupled and R4 is tuned to be critically coupled, the bandpass filter can have

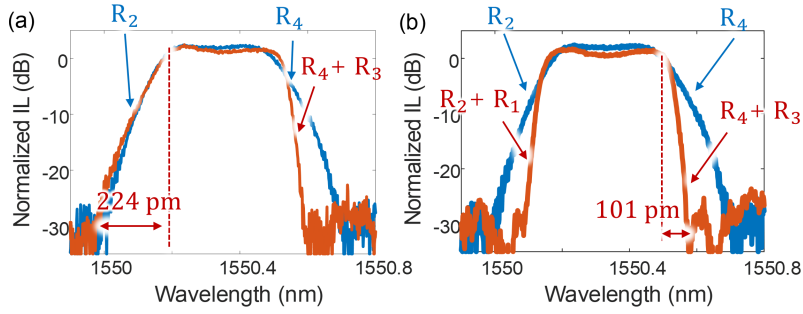

**Fig. S2.16** (a) R3 is added to improve the roll-off. (b) R1 and R3 are added to improve the roll-off.

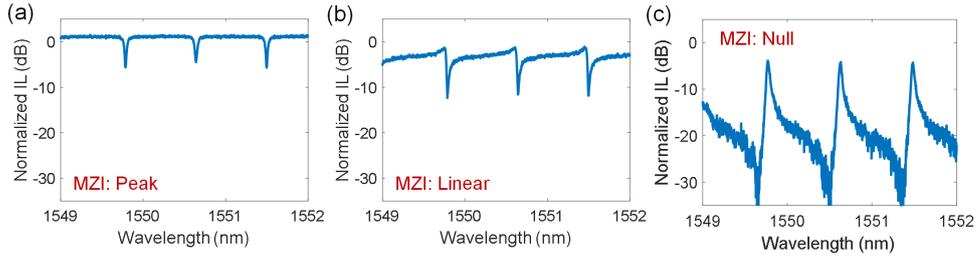

**Fig. S2.17** Response of a MZI loaded with a critical-coupled ring. (a) MZI biased at peak point; (b) MZI biased at linear point; (c) MZI biased at null point;.

a narrower (50 pm) bandwidth and a lower loss (3 dB extra loss). And if these two rings are both critically coupled, the filter passband can be as narrow as 29.5 pm with a loss of 7 dB.

The roll-off transition can also be improved with a higher-order filter. As can be seen in Fig. S2.16, extra cascaded ring filters (R1 and R3) can help to increase the all-pass filter order and reach a narrower roll-off transition (from 224 pm to 101 pm). Thus, if higher order filters are required, more ring filters can be added on both arms.

When the ring filters are critically coupled, the ring loss can no longer be neglected and the rings are no longer all-pass filters. If the splitter and combiner's $\kappa$ of the filter block is set to 0 or 1, the rings act as independent tunable band-stop filters, as shown in Fig. S2.12 (b)-(e). When the splitter and combiner's $\kappa$ of the filter block is set to 0.5, forming a good MZI, the responses of the MZI with a critically coupled ring are shown in Fig. S2.17. It can be seen that a bandpass filter with a Fano resonance is achieved when the MZI is biased at null point in Fig. S2.17(c). By tuning the ring phase, a tunable bandpass filter is achieved, as shown in Fig. S2.18(a). And if two rings are used, then we can get two pass bands which can be independently tuned, shown in Fig. S2.18(b). The 3 dB bandwidth of these passbands from R1 are around 35 pm, which is slightly larger than the 3 dB bandwidth of R1 (30.7 pm).

In conclusion, the filter block of our signal processor can in principle be configured for arbitrary magnitude responses (limited by the performance of fabricated circuits). We have shown measurement results with bandstop filters (with single rings), bandpass filters (Ring-loaded MZI with critical coupled rings), and flat-top Chebyshev type II type bandpass filters (Ring-loaded MZI with over coupled rings). Thanks to all the tap PDs, the system can be monitored and calibrated locally.

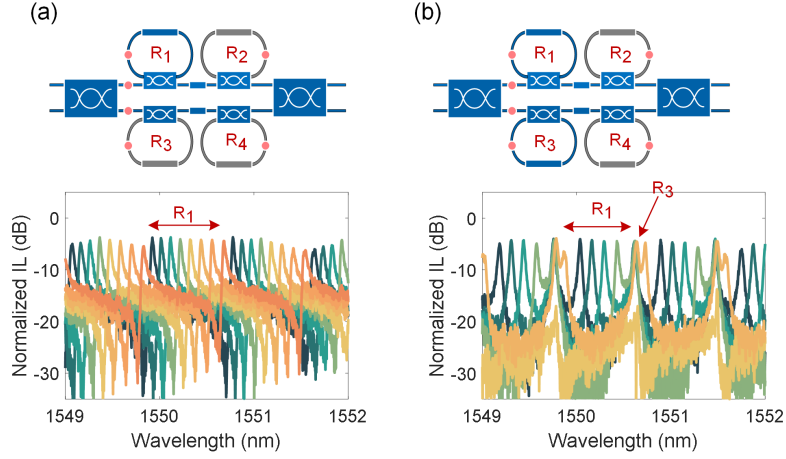

**Fig. S2.18** (a) A tunable bandpass filter; (b) An independently tunable two pass-band filter.

## S2.5 PD Block: High-speed photodiodes

There are two high-speed PDs in our signal processor for optical-to-electrical conversion. Due to the lack of transimpedance amplifiers (TIA), the PDs are directly wirebonded to the PCB and connected with RF connectors.

The measurement setup and the results are shown in Fig. S2.19. Here, we used an off-chip PD (Discovery Labbuddy DSC10H, 43 GHz, 0.54 A/W at 1550 nm) and an on-chip PD to measure the on-chip modulator's response with the same configuration, and the results are shown in Fig. S2.19(b). Then we can use the off-chip PD's response as a reference to get the on-chip PD's response. The differential result shows a 3 dB bandwidth of 25 GHz, as shown in Fig. S2.19(c). The PD design is listed as 50 GHz in the PDK document, thus the RF packaging (including the wire-bonding process) limits the PD's bandwidth.

As discussed in Section 1, the RF packaging is not ideal and there is a high crosstalk which the RF signal couples from input to output port directly. Comparison of the on-chip PD recovered signal and the electrical crosstalk is shown in Fig. S2.20(a), and the extracted PD response is shown in Fig. S2.20(b). As can be seen, beyond 25 GHz, the PD signal rapidly dropped to the level of the crosstalk . Thus, the packaged system can only work up to 25 GHz. Due to the high crosstalk, the PD recovered signal will be immersed in the crosstalk if only the on-chip laser is used. Thus, we used an off-chip optical amplifier (EDFA) and fed the pumped light back to the on-chip PD. If a better RF packaging and a proper TIA can be implemented, this extra optical amplification is not necessary.

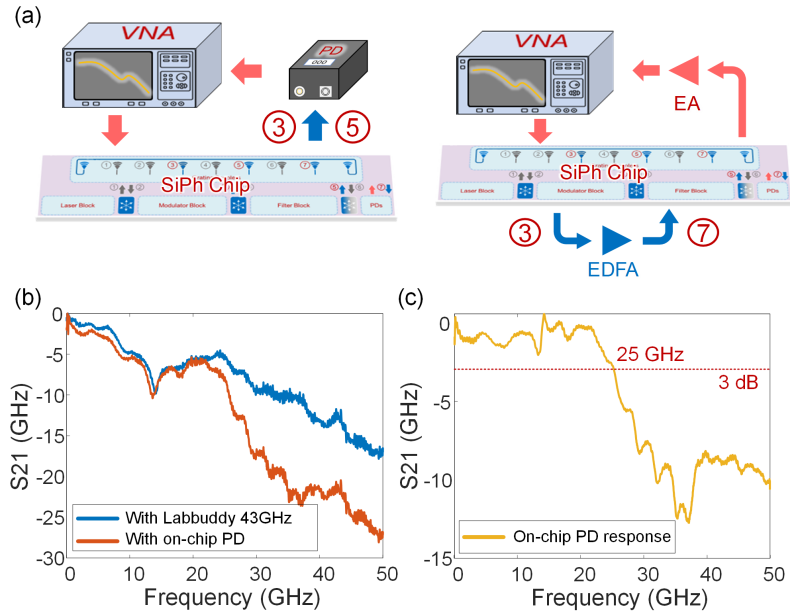

**Fig. S2.19** (a) RF responses measurement setups with an off-chip PD and with an on-chip PD; (b) The modulator block's RF responses with an off-chip PD and with an on-chip PD; (c) The on-chip PD response using the off-chip PD's response as a reference. VNA: Vector network analyzer. EDFA: Erbium-doped fiber amplifier. EA: Electronic amplifier.

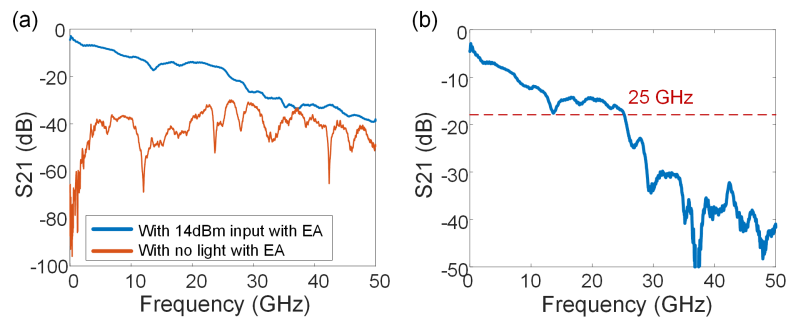

**Fig. S2.20** (a) A comparison of on-chip PD response and RF crosstalk; (b) The crosstalk-removed PD response.

17

# S3 Chip Applications

With the characterized and calibrated optical system, we demonstrated several applications with our signal processor.

## S3.1 Microwave Photonic Filter

One target application for our signal processor is microwave photonic filtering. Here we assume an input RF signal $V_{\text{in}}(t)$ as a board band arbitrary signal. To get the small signal modelling of the system, this input signal can be decomposed as a large offset signal and a small signal variance as:

$$V_{\text{in}}(t) = V_0 + v_{\text{in}}(t) \tag{S3.1}$$

where $V_0$ is the offset signal and $v_{\text{in}}(t)$ is the small signal. Then the modulated light phase can be expressed as:

$$\begin{aligned}\phi_m &= \frac{\pi}{V_\pi} V_{\text{in}}(t) \\ &= \frac{\pi}{V_\pi}(V_0 + v_{\text{in}}(t))\end{aligned} \tag{S3.2}$$

Instead of using Jacobi-Anger expansion in Eq. S2.22 , we can use Taylor expansion around $V_0$ here to get the linear transfer function of the system:

$$\begin{aligned}e^{-j\phi_m} &= e^{-j\frac{\pi}{V_\pi}V_{\text{in}}(t)} \\ &= (e^{-j\frac{\pi}{V_\pi}V_0} - j\frac{\pi}{V_\pi}e^{-j\frac{\pi}{V_\pi}V_0}v_{\text{in}}(t) - \frac{1}{2}(\frac{\pi}{V_\pi})^2 e^{-j\frac{\pi}{V_\pi}V_0} v_{\text{in}}^2(t) + ...) \\ &= (1 - j\frac{\pi}{V_\pi}v_{\text{in}}(t) - \frac{1}{2}(\frac{\pi}{V_\pi})^2 v_{\text{in}}^2(t) + ...)e^{-j(\frac{\pi}{V_\pi}V_0)}\end{aligned} \tag{S3.3}$$

Take it into Eq. S2.21, we can get the output field of the modulator block as:

$$\begin{aligned}E_{\text{out1}} &= [(1-\kappa)\alpha e^{-j\phi_m} - \kappa e^{-j\phi_s}]E_{\text{in1}} \\ &= (1-\kappa)\alpha(1 - j\frac{\pi}{V_\pi}v_{\text{in}}(t) - \frac{1}{2}(\frac{\pi}{V_\pi})^2 v_{\text{in}}^2(t)^2 + ...)e^{-j\frac{\pi}{V_\pi}V_0}e^{-j\omega_c t} \\ &\quad - \kappa e^{-j\phi_s}e^{-j\omega_c t}\end{aligned} \tag{S3.4}$$

for $\phi_s$ is defined as the phase difference of the two arm, it can cover the static shifts of $V_0$. If we just focus on the first order of sidelobes, the output field can be simplified as:

$$E_{\text{out1}} \sim [(1-\kappa)\alpha - \kappa e^{-j\phi_s}) + (1-\kappa)\alpha(-j\frac{\pi}{V_\pi}v_{\text{in}}(t)]e^{-j(\omega_c t)} \tag{S3.5}$$

If we assume the modulator phase modulation is linear, the input RF signal $v_{\text{in}}(t)$ can be expressed as a summation of different frequency components, as:

$$\begin{aligned}v_{\text{in}}(t) &= \sum^{\omega_s}(v_s \cos(\omega_s t + \phi_{\omega_s})) \\ &= \sum^{\omega_s}(v_s(\frac{e^{-j(\omega_s t + \phi_{\omega_s})} + e^{j(\omega_s t + \phi_{\omega_s})}}{2}))\end{aligned} \tag{S3.6}$$

Thus the modulated signal can be expressed as:

$$\begin{aligned}E_{\text{out1}} &= \sum^{\omega_s}[(1-\kappa)\alpha - \kappa e^{-j\phi_s})e^{-j\omega_c t} \\ &\quad - j(1-\kappa)\alpha\frac{\pi v_s}{2V_\pi}(e^{-j(\omega_s t + \phi_{\omega_s})} + e^{j(\omega_s t + \phi_{\omega_s})})e^{-j\omega_c t}] \\ &= \sum^{\omega_s}[M_0 e^{-j\phi_{M_0}}e^{-j\omega_c t} + jM_1 v_s(e^{-j(\omega_s t + \phi_{\omega_s})} + e^{-j(-\omega_s t - \phi_{\omega_s})})e^{-j\omega_c t}]\end{aligned} \tag{S3.7}$$

where now
$$M_0 e^{-j\phi_{M_0}} = (1-\kappa)\alpha - \kappa e^{-j\phi_s}$$
$$M_1 = -(1-\kappa)\alpha \frac{\pi}{2V_\pi} \quad (S3.8)$$

By tuning the tunable couplers' coupling ratio $\kappa$ and the phase shifter $\phi_s$ in the modulator block, we can arbitrarily set $M_0$, $\phi_{M_0}$, and $M_1$

For a broadband RF signal modulated light signal, the optical spectrum can be shown as Fig. S3.1(a). If the signal goes through an optical filter, the spectrum will be shaped in magnitude and phase as shown in Fig. S3.1(b). Here we set the transmission of the optical filter as::

- Lower sideband: $L(\omega_c - \omega_s)e^{-j\phi_L(\omega_c-\omega_s)}$
- Upper sideband: $U(\omega_c + \omega_s)e^{-j\phi_U(\omega_c+\omega_s)}$
- Central frequency: $C(\omega_c)e^{-j\phi_C(\omega_c)}$

The filtered optical signal can be expressed as:

$$E_{\text{filtered}} = \sum^{\omega_s}[M_0 C e^{-j(\phi_{M_0}+\phi_c)}e^{-j\omega_c t} + \\ jM_1 v_s(U e^{-j(\omega_s t+\phi_{\omega_s}+\phi_U)} + L e^{-j(-\omega_s t-\phi_{\omega_s}+\phi_L)})e^{-j\omega_c t}] \quad (S3.9)$$

Then the signal is fed into the PD, and the generated photocurrent is:

$$I_{\text{PD}} = \mathcal{R} E_{\text{filtered}} E^*_{\text{filtered}} \quad (S3.10)$$

where $\mathcal{R}$ is the responsibility of the PD. And $I_{\text{PD}}$ can be calculated as:

$$I_{\text{PD}} = \mathcal{R}\sum^{\omega_s}[M_0 C e^{-j(\phi_{M_0}+\phi_c)}e^{-j\omega_c t} + \\ jM_1 v_s(U e^{-j(\omega_s t+\phi_{\omega_s}+\phi_U)} + L e^{-j(-\omega_s t-\phi_{\omega_s}+\phi_L)})e^{-j\omega_c t}][M_0 C e^{j(\phi_{M_0}+\phi_c)}e^{j\omega_c t} + \\ jM_1 v_s(U e^{j(\omega_s t+\phi_{\omega_s}+\phi_U)} + L e^{j(-\omega_s t-\phi_{\omega_s}+\phi_L)})e^{j\omega_c t}] + \text{mixed signal} \\ = I_{\text{dc}} + I_{1^{st}} + I_{2^{nd}} + \text{mixed signal} \quad (S3.11)$$

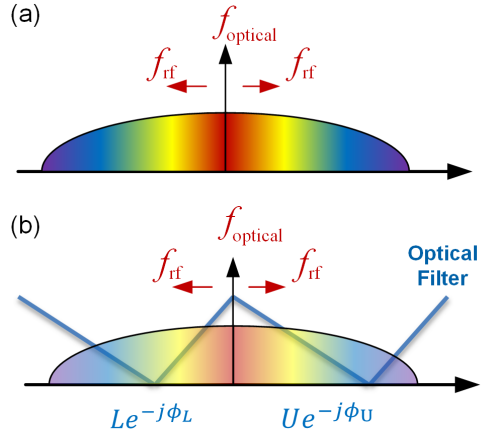

**Fig. S3.1** (a) Optical spectrum of a modulated light signal; (b) A modulated light signal with an optical filter.

where $I_{\text{dc}}$, $I_{1^{st}}$, $I_{2^{nd}}$ are the DC signal, first order signal (original frequency $\omega_s$), and second order harmonics (doubled frequency $2\omega_s$), respectively. Here again, if we just focus on the linear response $I_{1^{st}}$ of this microwave photonic filtering system and ignore the mixed signal, we can get that:

$$I_{1^{st}} = 2\mathcal{R} \sum^{\omega_s} M_1 M_0 C v_s [U \sin(\omega_s t + \phi_{\omega_s} + \phi_U - \phi_{M_0} - \phi_C) \\ + L \sin(-\omega_s t - \phi_{\omega_s} + \phi_L - \phi_{M_0} - \phi_C)] \quad (S3.12)$$

If we get the phase term out, it becomes:

$$I_{1^{st}} = 2\mathcal{R} M_1 M_0 C [\sum^{\omega_s} (U e^{-j(\phi_U - \phi_{M_0} - \phi_C - \pi/2)} v_s \cos(\omega_s t + \phi_{\omega_s}) - \\ L e^{-j(\phi_L - \phi_{M_0} - \phi_C - \pi/2)} v_s \cos(\omega_s t + \phi_{\omega_s}))] \\ = 2\mathcal{R} M_1 M_0 C [U e^{-j(\phi_U - \phi_{M_0} - \phi_C - \pi/2)} - L e^{-j(\phi_L - \phi_{M_0} - \phi_C - \pi/2)}] v_{\text{in}}(t) \quad (S3.13)$$

This shows the full expression of the linear frequency response, and it can be rewritten as:

$$I_{1^{st}} = |H_{\omega_s}| e^{-j\phi_{\omega_s}} v_{\text{in}} \quad (S3.14)$$

where

$$|H_{\omega_s}| = |2\mathcal{R} M_1 M_0 C \sqrt{U^2 + L^2 - 2UL \cos(\phi_U + \phi_L - 2(\phi_{M_0} + \phi_C))}| \\ \phi_{\omega_s} = \arctan(\frac{-U \cos(\phi_U - (\phi_{M_0} + \phi_c)) + L \cos(\phi_L - (\phi_{M_0} + \phi_c))}{U \sin(\phi_U - (\phi_{M_0} + \phi_c)) + L \sin(\phi_L - (\phi_{M_0} + \phi_c))}) \quad (S3.15)$$

and in which

$$M_0 e^{-j\phi_{M_0}} = (1-\kappa)\alpha - \kappa e^{-j\phi_s} \\ M_1 = -(1-\kappa)\alpha \frac{\pi}{2V_\pi} \quad (S3.16)$$

Thus, we got the full transfer function for the original frequency $\omega_s$:

$$H = 2\mathcal{R} M_1 M_0 C [U e^{-j(\phi_U - \phi_{M_0} - \phi_C - \pi/2)} - L e^{-j(\phi_L - \phi_{M_0} - \phi_C - \pi/2)}] \\ = |H_{\omega_s}| e^{-j\phi_{\omega_s}} \quad (S3.17)$$

In reality, the used phase modulator cannot have a perfect linear phase response, while the tuning of modulator's $\kappa$ and $\phi_s$ can linearize the phase modulation [2], and the tuning of the optical filter ($U$ and $L$) can compensate the imperfection. Thus, this system works with slightly tuned parameters when an imperfect phase modulator is used.

If we make an assumption that the lower sideband region of the optical transmission spectrum ($L$) and the upper sideband region ($U$) are fully independent (which is not true for continuous broadband spectrum transmissions, but is a valid assumption for microwave signals modulated on a carrier frequency), the microwave photonic filter design can also use allpass filters to define a filter function similar to the optical filter design discussed in Section S2.4.. The "splitter" and "combiner" in this "interferometer" are implemented by the modulator and PD, which can be regarded as perfect 50:50 coupling ratio without any insertion loss, and the all-pass filter can still be ring resonators, and then it can also be regarded as a difference of two all-pass filters, as shown in Eq. S3.17. Because the ring resonators will both work in the upper and lower sidebands, the design method can only target at half of the FSR of the ring resonators (in our case, 53.5 GHz), and in the other half the response shape will be the image of the first one.

Here we show some demonstrations for the RF filter design. If the reconfigurable modulator is set as a phase modulator and the filter block is set as a single notch filter (a critical coupled ring resonator), we can achieve a single bandpass filter in the RF domain, as shown

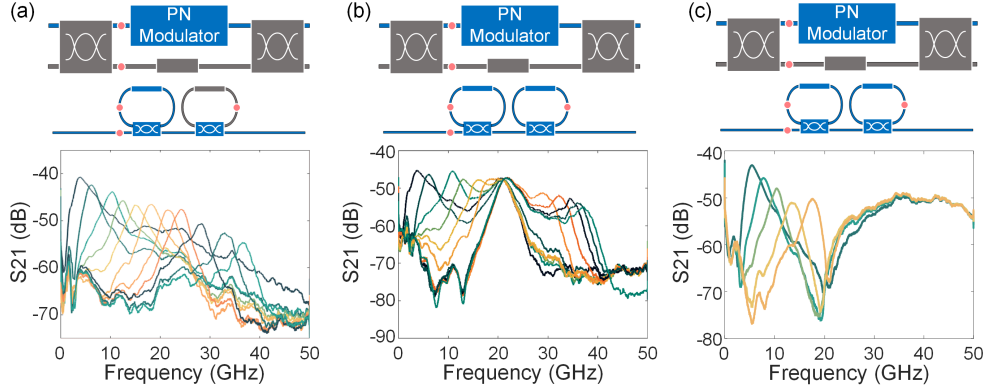

**Fig. S3.2** (a) Microwave filter: single tunable bandpass filter; (b) Microwave filter: double tunable bandpass filter; (c) Microwave filter: tunable one-pass one-stop band filter.

in Fig. S3.2(a). If two rings are used, a double bandpass filter or one-pass one-stop band filter can be realized, where the difference is that the second ring is under/critically coupled (shown in Fig. 3(b)), or overcoupled (shown in Fig. 3(c)). This modulator configuration is set for the $\kappa = 0$, thus $\phi_{A_0} = 0$ in Eq. S2.24, and these RF filtering results fully match the optical filtering response when single ring or double rings used in an MZI biasing at null point (Fig. S2.18).

If the ring resonators are over coupled, they act as all-pass filters for the RF signal loaded light carriers. By tuning the modulator block and the cascaded rings (R1 and R2), we can get tunable low-pass filters (LPF), high-pass filters (HPF), and bandpass filters (BPF), as shown in Fig. S3.3. Fig. S3.3(a), (d), and (h) show the measured S21 responses of the tunable filters, but these responses are enveloped by the response of the original modulator and PD response. Thus we constructed an "all pass" reference which combines the widest HPF and the widest LPF, as shown in Fig. S3.3(g), and all other curves are calibrated with the reference curve, which are shown in Fig. S3.3(b) and (e). These responses more clearly show the high-pass and low-pass behaviors. The curves are further normalized by removing the insertion loss in Fig. S3.3(c) and (f). Because the responses below $-70$ dB are close to the noise floor at high frequency, the low-pass filter responses are not accurate at the high frequency points, thus we truncate the responses in Fig. S3.3(c) to 30 GHz, to make it more clear. Fig. S3.3(h) shows a tunable BPF response. Because of the large roll-off of the optical ring resonator, these band-pass filters show higher losses than previous filters. The normalized BPF responses are shown in Fig. S3.3(i). The high-pass, low-pass and band-pass filters are realized with the same mechanism as the optical band-pass filters discussed in Fig. S2.15, while the RF filters are in a narrower frequency/wavelength range (30 GHz to 240 pm range at 1550 nm).

The full optical filter block can also be used to form microwave photonic filters. The final RF responses can be fully predicted with Eq. S3.5, but its filter synthesis will be hard. Fig. S3.4 shows a four pass-band filter where these four bands can be independently tuned and suppressed.

One important application of the microwave photonic filter is an RF equalizer. Here we implement it for our intensity modulator with PDs. The results can be seen in Fig. S3.5. The curves in Fig. S3.5 are the raw data from VNA and not calibrated with other data, which means that all cables, bonding wires and other components' response are included. It shows

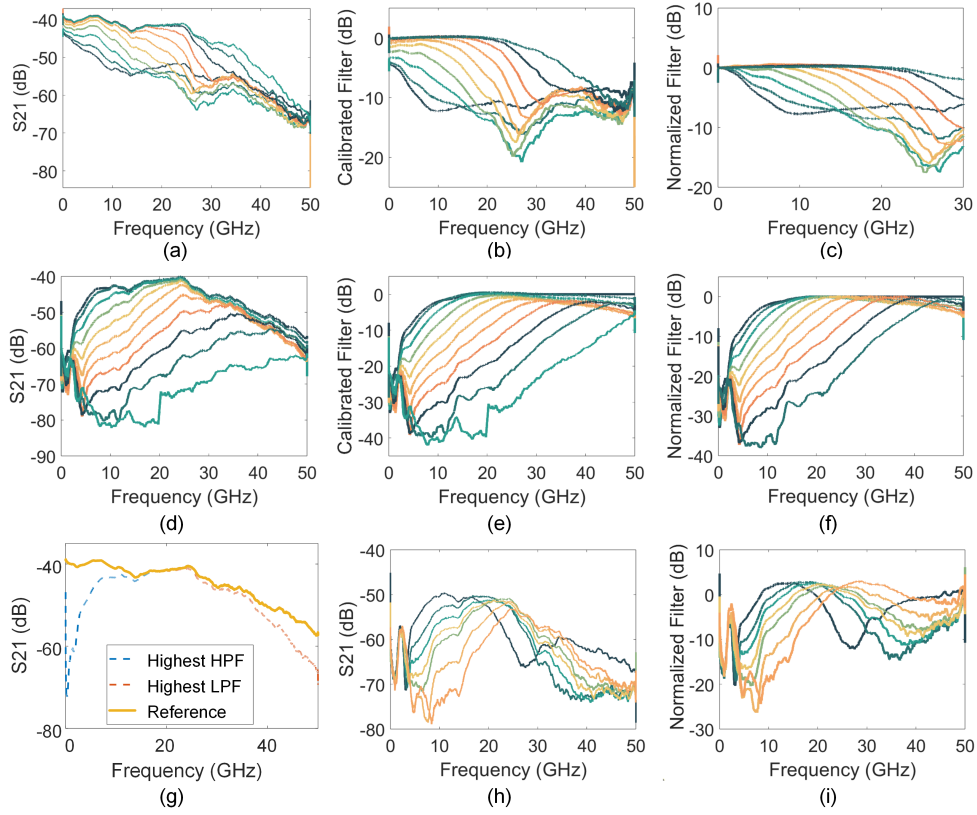

**Fig. S3.3** (a) Measured S21 response of a tunable low-pass filter (LPF); (b) Using the highest one as a reference, calibrated filter response of the tunable LFP; (c) Normalized filter response of the tunable LFP; (d) Measured S21 response of a tunable high-pass filter (HPF); (e) Using the highest one as a reference, calibrated filter response of the tunable HPF; (f) Normalized filter response of the tunable HPF; (g)A "all pass" reference curve generated from the highest HPF and highest LFP, to calibrate the remaining filters. (h) Measured S21 response of a tunable band-pass filter (BPF); (i) Normalized filter response of the tunable BPF.

that the microwave filter based equalizer can flatten the response of the whole system ith an off-chip PD up to 26 GHz (Fig. S3.5(a)) and with an on-chip PD up to 25 GHz (Fig. S3.5(b)). The equalized curve in Fig. S3.5(a) is partially higher than unequalized one is because the used EDFA provides a constant output power, and the optical sidelobes are being amplified more when the optical carrier is suppressed.

## S3.2 Frequency Multiplier

Another application for our signal processor is the frequency multiplier. Here we used the modulator block to achieve frequency doubling with high original signal extinction. It is based on the carrier-suppressed double-sideband modulation scheme with the reconfigurable modulator, the configuration of which is shown in Fig. S3.6 [3]. The main advantage of the

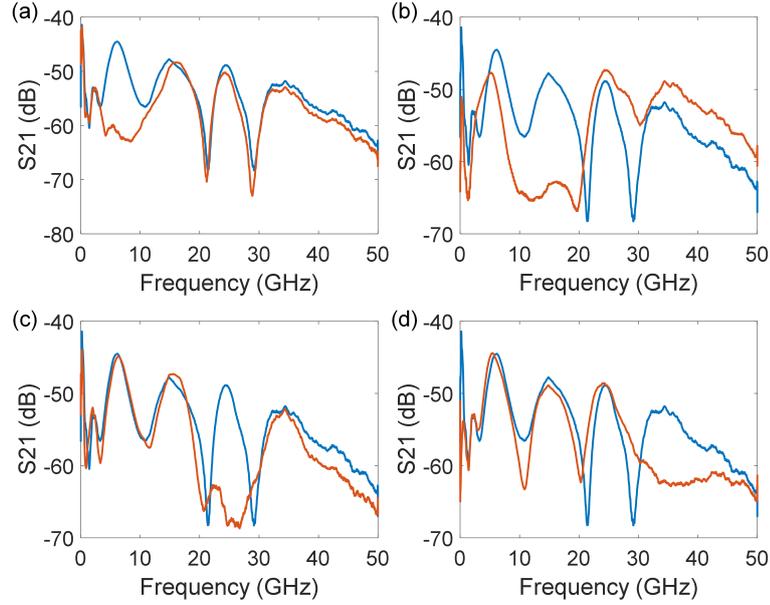

**Fig. S3.4** A tunable four band-pass filter and its first (a), second (b), third(c), and forth(d) pass-band suppressed filter.

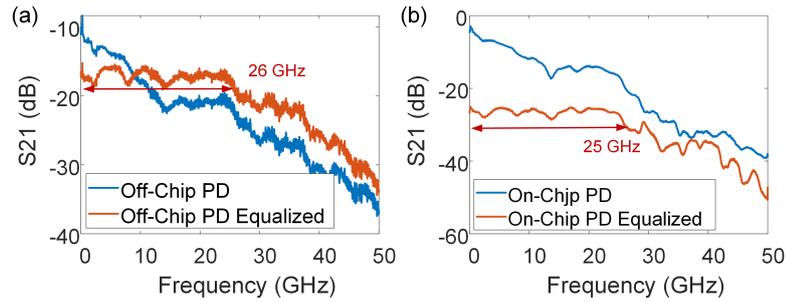

**Fig. S3.5** (a) Equalized transmission spectrum with an off-chip PD; (b) Equalized transmission spectrum with an on-chip PD

system is that the tunable splitter used in the modulator block can be tuned to compensate the extra loss of the used modulator, and then reach a high carrier extinction.

Demonstrations were implemented with our signal processor, as shown in Fig. S3.7. Here we used an RF source (R&S SMA40) as an RF signal generator, The generated RF signal will be modulated on the laser carrier (from on-chip laser block) by the modulator block, and then the modulated light signal will be coupled to a PD. The recovered RF signal was fed into a spectrum analyzer. The frequency-doubling results are shown in Fig. S3.8. In this part, the extinction ratio is calculated by the second harmonic signal power to the original frequency signal power. As can be seen, normal intensity modulation can keep the second harmonic signal to be 40 dB lower than the original frequency, while if the modulator is biased at null

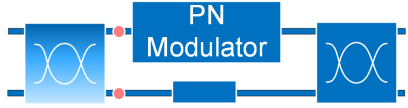

**Fig. S3.6** Modulator block configuration for frequency doubling, where more light is guided to the PN modulator for compensating its insertion loss.

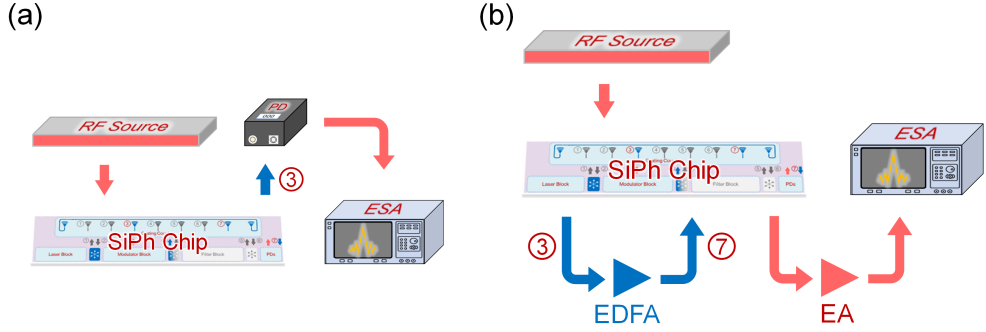

**Fig. S3.7** (a) A demonstrator for frequency doubling with an off-chip PD; (b) A demonstrator for frequency doubling with an on-chip PD. SiPh Chip: Silicon photonic chip. EDFA: Erbium-doped fiber amplifier. EA: Electronic amplifier. ESA: Electrical spectrum analyzer.

point, the second harmonic signal can be 40 dB higher than the original one. The extinction ratio's drop is mainly due to the system RF loss for higher frequency.

We also did the same measurement with an on-chip PD (Fig. S3.7(b)), and the results are shown in Fig. S3.9. As it is shown, this frequency doubling can also work for 20 GHz, but the extinction ratio degenerates a lot at the higher frequencies. This is due to the high RF crosstalk at high frequencies, where the first-order signal is not from the optical signal but is from the direct coupling crosstalk, which can be seen in Fig. S3.9(b). One thing should be noted is that the first-order signal power is lower than the crosstalk below 10 GHz, which may because of the RF signal destructive interference between the light generated RF signal with the crosstalk RF signal.

### S3.3 Opto-electronic Oscillator

Our signal processor can not only generate optical signals with the on-chip laser block, but also can form an opto-electronic oscillator for RF signal generation. We also did the tests with an off-chip PD or an on-chip PD, and their schematics are shown in Fig. S3.10. Here, the on-chip laser, modulator, optical filter blocks, as well as PDs formed a tunable bandpass microwave photonic filter, as shown in Fig. S3.2(a). The modulated RF signal is loaded on the optical link, filtered, amplified by an electronic amplifier and then fed back into the modulator as an RF source. The whole system thus becomes an RF loop, and it will start oscillation if the net gain is 0 dB or higher. By tuning the pass-bands of the microwave photonic filter, as discussed in the Section S3.1, we can tune the oscillation RF frequency and achieve a tunable RF source. A length of extra fiber was added to the system, which extends the entire cavity to achieve a lower RF phase noise.

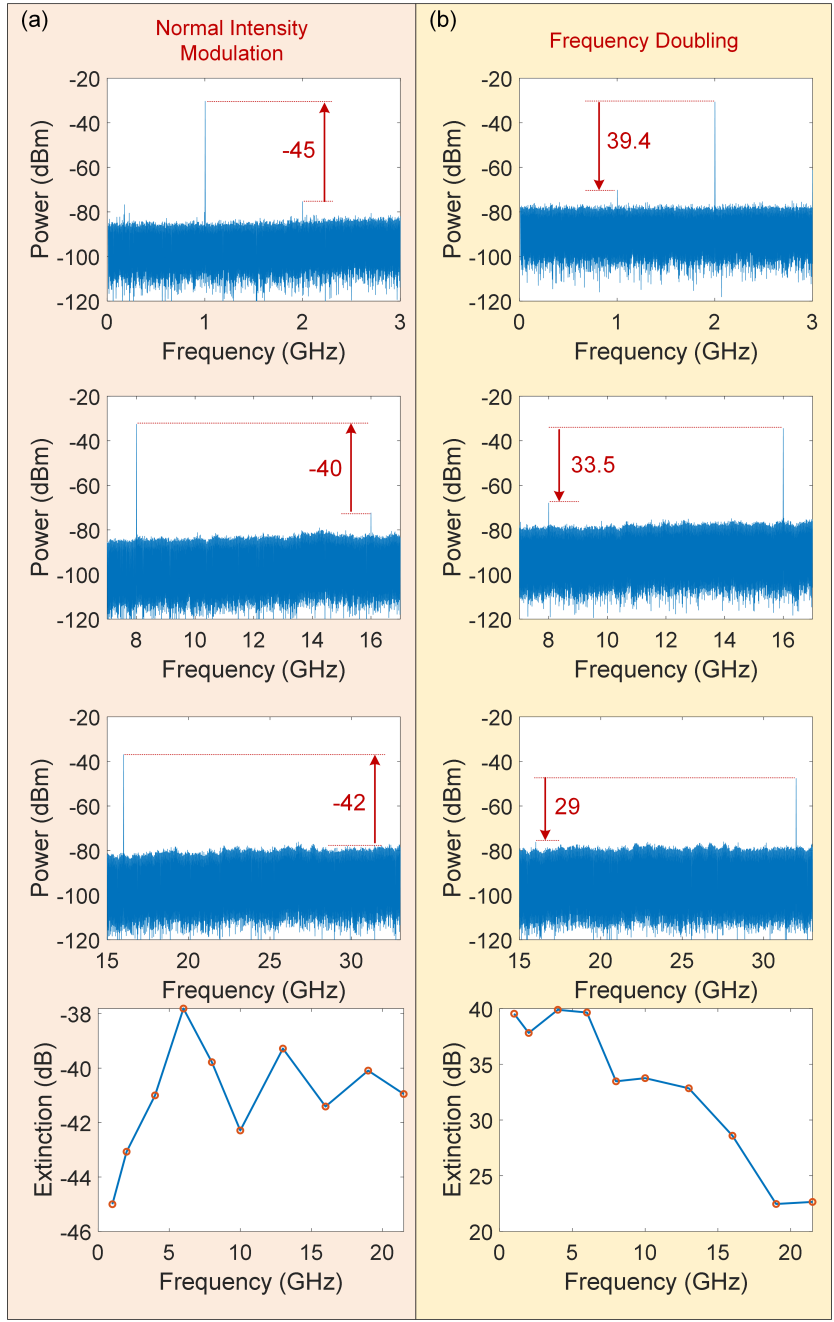

**Fig. S3.8** Frequency multiplying results with an off-chip PD. (a) Normal intensity modulation when the modulator MZI is biased at quadrature point; (b) Frequency doubling when the modulator MZI is biased at null point.

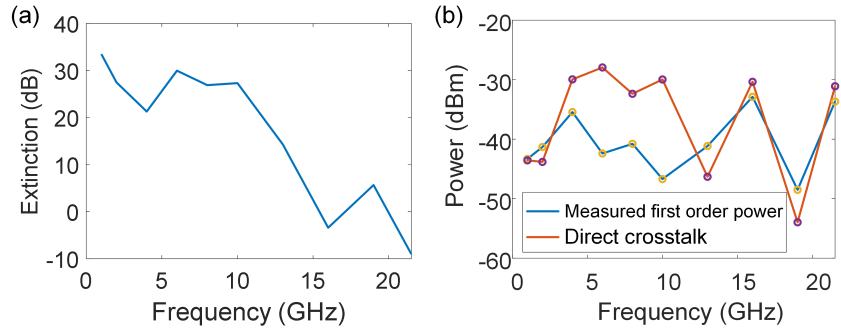

**Fig. S3.9** Frequency multiplying results with an on-chip PD. (a) Frequency doubling when the modulator MZI is biased at null point. (c) Measured first order signal power and directly coupled crosstalk power.

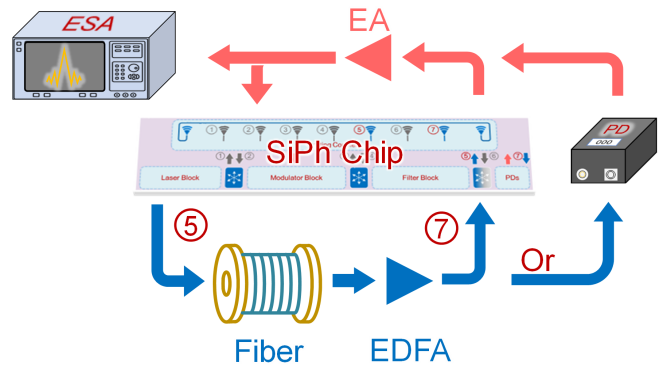

**Fig. S3.10** An opto-electronic oscillator demonstrator with the single chip signal processor. SiPh Chip: Silicon photonic chip. EDFA: Erbium-doped fiber amplifier. EA: Electronic amplifier. ESA: Electrical spectrum analyzer.

The RF generation results with the off-chip PD are shown in Fig. S3.11. Here, the extra fiber added is 500 m long. The spectrum of generated RF signals is shown in Fig. S3.11(a) and (b). As can be seen, the signal frequency can be tuned from 3.8 GHz to 24.1 GHz, and higher frequencies can be reachable, but limited by the used ESA (Anritsu MS2840A, 26 GHz). The zoom-in spectrum shows that the mode spacing of this oscillator is around 400 kHz and the sidelobe extinction is 32 dB. Fig. S3.11(c) and (d) show the phase noise of the generated signal. It can be noted that the strength of the phase noise is almost constant for all the generated signals.

Demonstrators with on-chip PDs are also implemented, and the results are shown in Fig. S3.12. Here, the extra fiber added is 20 m long, and the oscillator cannot keep a single mode oscillation with a longer fiber. The spectrum of the generated RF signals are shown in Fig. S3.12(a) and (b). As can be seen, the signal frequency can be tuned from 3.3 GHz to 11.6 GHz. The zoom-in spectrum shows that the mode spacing of this oscillator is around 4 MHz and the sidelobe extinction is 18.7 dB. Fig. S3.11(c) and (d) show the phase noise of

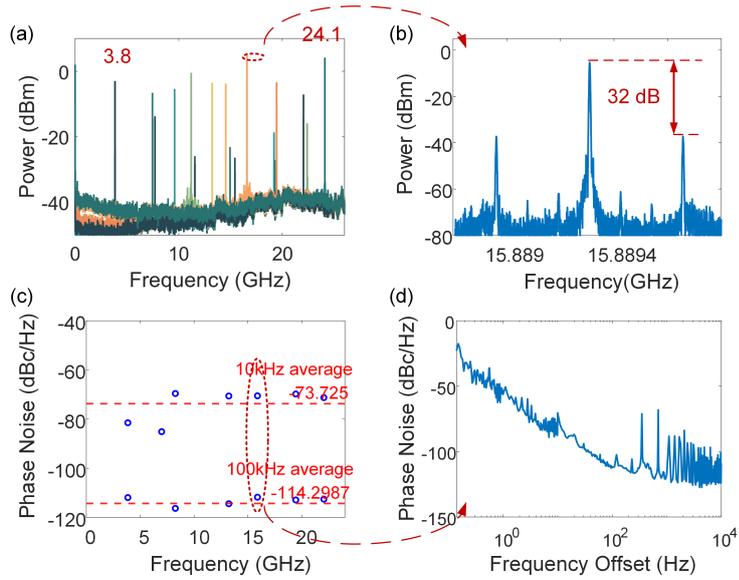

**Fig. S3.11** (a) Tunable RF signal generation from 3.8 GHz to 24.1 GHz with an off-chip PD; (b) The zoom-in spectrum of generated 15.8 GHz signal; (c) Extracted phase noises of the generated RF signals; (d) Phase noise results of the generated 15.8 GHz signal.

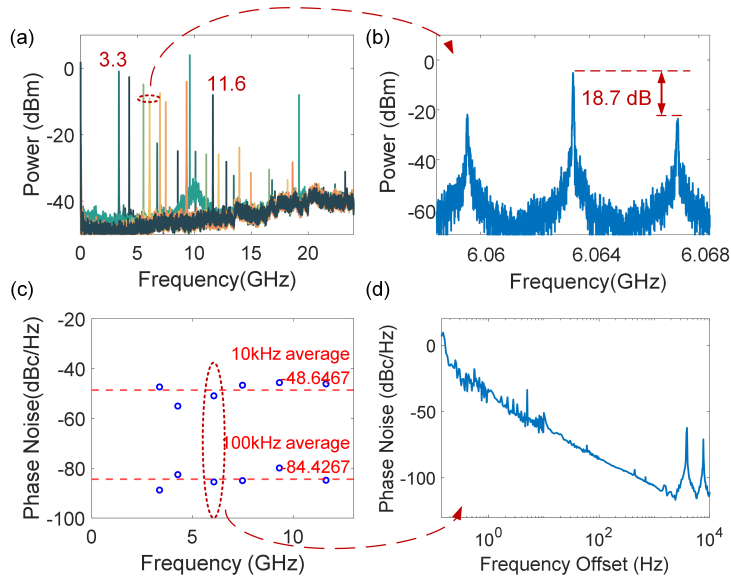

**Fig. S3.12** (a) Tunable RF signal generation from 3.3 GHz to 11.6 GHz with an on-chip PD; (b) The zoom-in spectrum of generated 6.06 GHz signal; (c) Extracted phase noises of the generated RF signals; (d) Phase noise results of the generated 6.06 GHz signal.

27

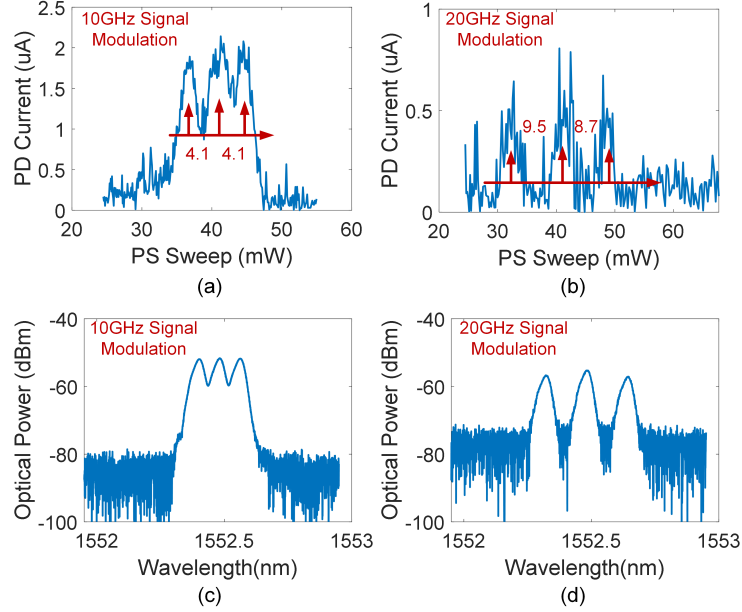

**Fig. S3.13** A demonstrator as an optical wavelength meter or an RF frequency meter. (a) and (c) The light carrier with 10 GHz RF signal modulated sidelobes; (b) and (d) The light carrier with 20 GHz RF signal modulated sidelobes. (a) and (b) is measured with current meter from tap PD's photocurrent. (c) and (d) is measured with an optical spectrum analyzer.

the generated signal. From this we can see that with the on-chip PD, the oscillator has a higher noise than with an off-chip PD ($-97$ dBc/Hz at 100 kHz offset with 20 m fiber), which is due to the RF packaging crosstalk. And crosstalk also makes the oscillator unstable to work higher than 11.6 GHz.

## S3.4 Other applications

Apart from the applications discussed above, our signal processor has more possibilities. Here we show that it can work as an optical wavelength meter or RF frequency meter.

The optical wavelength meter or the RF frequency meter is based on the ring resonators in the filter block. The optical power in the ring can be read with the tap PD. Then if we sweep the phase shifter in the ring, we can know the power distribution along the wavelength or frequency within one FSR of the ring. The results are shown in Fig. S3.13. To avoid that the main lobe fully covers the sidelobes, the modulator block is configured to have a carrier-suppressed (not fully) modulation. From Fig. S3.13 (a) and (b) we can see that the optical carriers and modulation introduced sidelobes can be easily recognized from the PD's readout. Compared with the measurement results from an optical spectrum analyzer Fig. S3.13 (c) and (d), the overall shape are revealed in Fig. S3.13 (a) and (b). The frequency measured (a) 7.8 GHz and (b) 17.1 GHz are not accurate, but a more dedicated calibration could overcome this limitation. The spectral resolution of this optical wavelength meter or RF frequency meter is determined by the 3 dB bandwidth of the used ring resonator. A ring resonator with a higher $Q$ factor can therefore reach a higher spectrum resolution.

# S4 Conclusion

In conclusion, the single-chip signal processor presented in this study serves as a versatile reconfigurable filter for both optical signals (S2.4) and RF signals (S3.1), while also providing optimized electro-optical (E/O) (S2.3) and opto-electrical (O/E) conversions (S2.5). The implementation of tunable laser sources through micro-transfer-printing allows for an impressive tuning range of up to 90 nm (S2.1). The system's flexibility is enhanced by using multiple monitor detectors, enabling local configuration. Furthermore, the processor exhibits remarkable capabilities in RF signal generation (S3.3) and RF frequency doubling (S3.2). Overall, this demonstration offers a promising and comprehensive approach to generating, distributing, and processing both optical and RF signals, with potential applications in data centers, wireless and satellite communication, and various optical and microwave fields. The seamless integration of functionalities on a single chip represents a significant advancement in signal processing technology.

[8] Wang M, Wang M, Ribero A, et al (2020) Tolerant, broadband tunable 2 &#x00D7; 2 coupler circuit. Optics Express, Vol 28, Issue 4, pp 5555-5566 28(4):5555–5566. https://doi.org/10.1364/OE.384018, URL https://opg.optica.org/viewmedia.cfm?uri=oe-28-4-5555&seq=0&html=truehttps://opg.optica.org/abstract.cfm?uri=oe-28-4-5555https://opg.optica.org/oe/abstract.cfm?uri=oe-28-4-5555

[9] Willson AN, Orchard HJ (1995) Insights into Digital Filters Made as the Sum of Two Allpass Functions. IEEE Transactions on Circuits and Systems I: Fundamental Theory and Applications 42(3):129–137. https://doi.org/10.1109/81.376877



\end{document}